\documentclass[twocolumn]{aastex63}
\submitjournal{AJ}
\received{November 19, 2021}
\usepackage{hyperref}
\usepackage[colorinlistoftodos]{todonotes}
\usepackage{amsmath}
\shorttitle{The Depth Metric}

\begin{document}
\title{Into the Depths: a new activity metric for high-precision radial velocity measurements based on line depth variations}

\correspondingauthor{Jared Siegel}
\email{siegeljc@uchicago.edu}

\author[0000-0002-9337-0902]{Jared C. Siegel}
\affiliation{Department of Astronomy and Astrophysics, University of Chicago, Chicago, IL 60637, USA}

\author[0000-0003-3856-3143]{Ryan A. Rubenzahl}
\altaffiliation{NSF Graduate Research Fellow.}
\affiliation{Department of Astronomy, California Institute of Technology, Pasadena, CA 91125, USA}

\author[0000-0003-1312-9391]{Samuel Halverson}
\affiliation{Jet Propulsion Laboratory, California Institute of Technology, 4800 Oak Grove Drive, Pasadena, California 91109}

\author[0000-0001-8638-0320]{Andrew W. Howard}
\affiliation{Department of Astronomy, California Institute of Technology, Pasadena, CA 91125, USA}

\begin{abstract}
The discovery and characterization of extrasolar planets using radial velocity (RV) measurements is limited by noise sources from the surfaces of host stars. Current techniques to suppress stellar magnetic activity rely on decorrelation using an activity indicator (e.g., strength of the Ca II lines, width of the cross-correlation function, broadband photometry) or measurement of the RVs using only a subset of spectral lines that have been shown to be insensitive to activity. Here, we combine the above techniques by constructing a high signal-to-noise activity indicator, the depth metric $\mathcal{D}(t)$, from the most activity-sensitive spectral lines using the ``line-by-line'' method of \cite{Dumusque2018}. Analogous to photometric decorrelation of RVs or Gaussian progress regression modeling of activity indices, time series modeling of $\mathcal{D}(t)$ reduces the amplitude of magnetic activity in RV measurements; in an $\alpha$CenB RV time series from HARPS, the RV RMS was reduced from 2.67 to 1.02 m s$^{-1}$. $\mathcal{D}(t)$ modeling enabled us to characterize injected planetary signals as small as 1 m s$^{-1}$. In terms of noise reduction and injected signal recovery, $\mathcal{D}(t)$ modeling outperforms activity mitigation via the selection of activity-insensitive spectral lines. For Sun-like stars with activity signals on the m~s$^{-1}$ level, the depth metric independently tracks rotationally modulated and multiyear stellar activity with a level of quality similar to that of the FWHM of the CCF and log$R^{\prime}_{HK}$. The depth metric and its elaborations will be a powerful tool in the mitigation of stellar magnetic activity, particularly as a means of connecting stellar activity to physical processes within host stars.
\end{abstract}


\section{Introduction}
\label{sec:intro}

The noise floor for the radial velocity (RV) detection of extrasolar planets is set by a combination of three sources: photon-limited uncertainties ($\sigma_{\rm photon}$), systematic errors in the instrument ($\sigma_{\rm sys}$), and stellar activity  ($\sigma_{\rm activity}$). Unlike $\sigma_{\rm photon}$, which depends on the signal-to-noise ratio of the spectrum, $\sigma_{\rm sys}$ and $\sigma_{\rm activity}$ both can have structured temporal dependence. For an overview of the various physical processes that give rise to RV noise, we refer the interested reader to the introduction section of \cite{deBeurs2020}. With improved resolution, calibration, and stabilization, next-generation instruments, such as the Keck Planet Finder~\citep[KPF;][]{Gibson2016}, ESPRESSO \citep{Megevand2014}, EXPRES \citep{Jurgenson2016}, and NEID \citep{Schwab2016}, are significantly lowering the $\sigma_{\rm sys}$ noise floor; KPF, a high-precision Doppler spectrometer scheduled for first light in the summer of 2022 on the Keck I telescope, is expected to achieve $\sigma_{\rm sys}=30$~cm~s$^{-1}$ and will achieve $\sigma_{\rm photon}=30$~cm~s$^{-1}$ on a $V= 10.9$ G2~star in only 30~min. While these advancements are significant, stellar activity on the m s$^{-1}$ level dominates over these small errors and limits the RV detections of small planets. We investigated novel stellar activity mitigation methods, with the goal of lowering the $\sigma_{\rm activity}$ noise floor in anticipation of KPF.

RV studies typically employ one of three methods of RV measurement: cross-correlation using a synthetic template \citep{Baranne1996}, maximum likelihood estimation to a reference spectrum \citep{Anglada2012}, or forward modeling \citep{Butler1996}. However, some spectral lines are more sensitive to systematic instrumental errors \citep{Dumusque2015} and others are more sensitive to stellar magnetic activity \citep{Thompson2017,Dumusque2018,Wise2018,Cretignier2020}. These three methods introduce noise into the derived RVs by treating all lines equally. 

Stellar activity and instrumental noise are often statistically accounted for through the addition of a time-independent white-noise jitter term to the RV uncertainties \citep[e.g., ][]{Wright2005}, or detrended with an activity model, such as linear decorrelation and Gaussian process (GP) regression \citep{Haywood2014,Rajpaul2015,Dai2017,Ahrer2021}. Activity models, whether parametric or nonparametric, commonly rely on a handful of spectral lines, such as the Ca II H\&K lines (log$R^{\prime}_{ HK}$) and H$\alpha$, or diagnostics from the cross-correlation function (CCF), such as the full width at half maximum (FWHM) and the bisector inverse slope (BIS), to track stellar activity \citep{Saar1998,Kurster2003, Queloz2009}. Alternatively, \cite{Aigrain2012} used a spot model to track stellar activity via the star's photometric light curve. This method, known as FF$^{\prime}$, has more recently been modified to use activity indicators in place of photometric flux \citep{Giguere2016}.

\cite{Dumusque2018} developed a novel approach to mitigating stellar noise by considering the Doppler shift of each spectral line independently; for HARPS spectra, ``line-by-line" RV measurement was shown to reach the same precision as standard techniques. \cite{Dumusque2018} also introduced the ``line selection method," in which only spectral lines shown to be activity-insensitive are considered for the RV measurement; the line selection method achieves a factor of $\sim$1.5 reduction in the $\alpha$CenB stellar noise. In a follow-up study, \cite{Cretignier2020} found that the effect of stellar activity on a given line is inversely proportional to the line's depth (a proxy for physical depth inside the stellar atmosphere). \cite{Cretignier2020} then demonstrated that using the difference between the RVs of deep and shallow spectral lines as an activity proxy (hereafter referred to as the ``formation depth method") yields better than a factor-of-two reduction in $\alpha$CenB's stellar activity amplitude.

In this paper, we develop a combination of the above line-by-line and detrending techniques into the depth metric method. Prior studies have demonstrated that, for a Sun-like star, more than 1,000 spectral lines are highly correlated with stellar activity \citep{Dumusque2018,Cretignier2020}. We combine a subpopulation of spectral lines shown to be activity-sensitive into a high signal-to-noise activity metric. Motivated by \cite{Wise2018}, we are particularly interested in the relationship between observation-to-observation variations in line depth and stellar activity; variable line depth has the benefit of being available at all wavelengths, is computationally inexpensive, is translation-invariant, and is a largely unexplored parameter space. Our depth metric can be modeled using time series activity mitigation techniques. The co-adding of orders of magnitude more lines than traditional indices has the potential to greatly improve modeling results \citep{Giguere2016}. To demonstrate the diagnostic power of the depth metric, we conducted a comparative study between linear decorrelation, FF$^{\prime}$, and GP models based on either the depth metric, log$R^{\prime}_{ HK}$, or the FWHM of the CCF; we also considered the line selection and formation depth methods. 

This paper is organized as follows. In Section \ref{sec:obs}, we describe the HARPS observations used in our study. Section \ref{sec:methods} summarizes our implementation of the line-by-line analysis technique of \cite{Dumusque2018} and develops our activity index, the depth metric. Then in Section \ref{sec:mitigation}, we describe the activity mitigation methods considered in our study, including the line selection method of \cite{Dumusque2018}, the formation depth method of \cite{Cretignier2020}, our modified FF$^{\prime}$ model, and our GP models. We applied these activity models to HARPS observations of $\alpha$CenB and HD~13808 in Section \ref{sec:model_comparison} and summarize our conclusions in Section \ref{sec:concl}.

\section{Observations  \label{sec:obs}}
We considered two stars observed with the HARPS spectrograph: $\alpha$CenB (HD~128621) and HD~13808. With high observation cadence and significant activity signals, these targets offer a valuable test bed for stellar activity mitigation methods. $\alpha$CenB presents a clear rotationally modulated stellar activity signal and lacks a significant planetary signal, although several candidates have been proposed \citep{Dumusque2012, Demory2015}. Between 2008 February and 2011 July, $\alpha$CenB was observed with HARPS between $1$ and $187$ times a night (median of 35). This campaign was designed to capture both long- and short-term stellar activity variations in $\alpha$CenB's RVs \citep{Dumusque2011}. As part of a sample of bright stars with relatively low levels of RV scatter, HD~13808 has been observed with HARPS since 2003 using exposure times greater than 15~minutes \citep{Ahrer2021}. HD~13808 exhibits a multiyear stellar magnetic activity cycle and hosts two recently confirmed Neptune-mass planets, allowing for a test of our mitigation methods' effects on planet characterization.

We correct each observation for known systematics following the prescriptions in \cite{Dumusque2018} and \cite{Cretignier2020}. Here, we briefly describe the corrections. We first divided the raw stellar spectrum by the blaze of the instrument. The resulting pixel fluxes were then continuum normalized via a rolling maximum, with a 10~$\rm \AA$ window size. To account for spectra with outlier continua, for each order index between 30 and 71 we fit the continuum (in the cross-dispersion direction) with a linear function. For a given target, we then rejected all spectra that had a single order's slope $>3\sigma$ from the mean slope of that order. Each spectrum's wavelength solution was reconstructed from the 288 polynomial coefficients stored in the spectrum's header. The wavelength solution was then Doppler shifted to correct for the barycentric Earth RV, which we adopted from the HARPS data reduction software (DRS), and was also Doppler shifted in accordance with the DRS-reported instrumental drift.

For $\alpha$CenB, we used the 2010 HARPS observations, over which the canonical activity indices (log$R^{\prime}_{HK}$, H$\alpha$, FWHM, and BIS) all present quasi-sinusoidal oscillations with periods near the $\sim$37~day rotation period \citep{DeWarf2010}. After rejecting all observations with an airmass greater than 1.5, we fit the HARPS reported RVs (stored in the header of each spectrum) using a quadratic polynomial plus a sinusoid as a function of time; the sinusoid component of the RV model tracks stellar activity, while the quadratic component accounts for long-term variations induced by $\alpha$CenA. We then performed 3$\sigma$ clipping on the DRS RVs with respect to our polynomial-sinusoid model for five iterations. The filtering process retained 908/1014 observations spanning 80 days. 

Following \cite{Cretignier2020}, we corrected for the long-term RV signal induced by $\alpha$CenB's binary companion by subtracting the quadratic component of the above model from the RVs. Via a least-squares regression, we found a sinusoid period of 35.9~days, nearly identical to that reported by \cite{Cretignier2020}. Unless otherwise stated, these corrected RVs are the ones used in our analyses.

\begin{figure*}[t]
\gridline{\fig{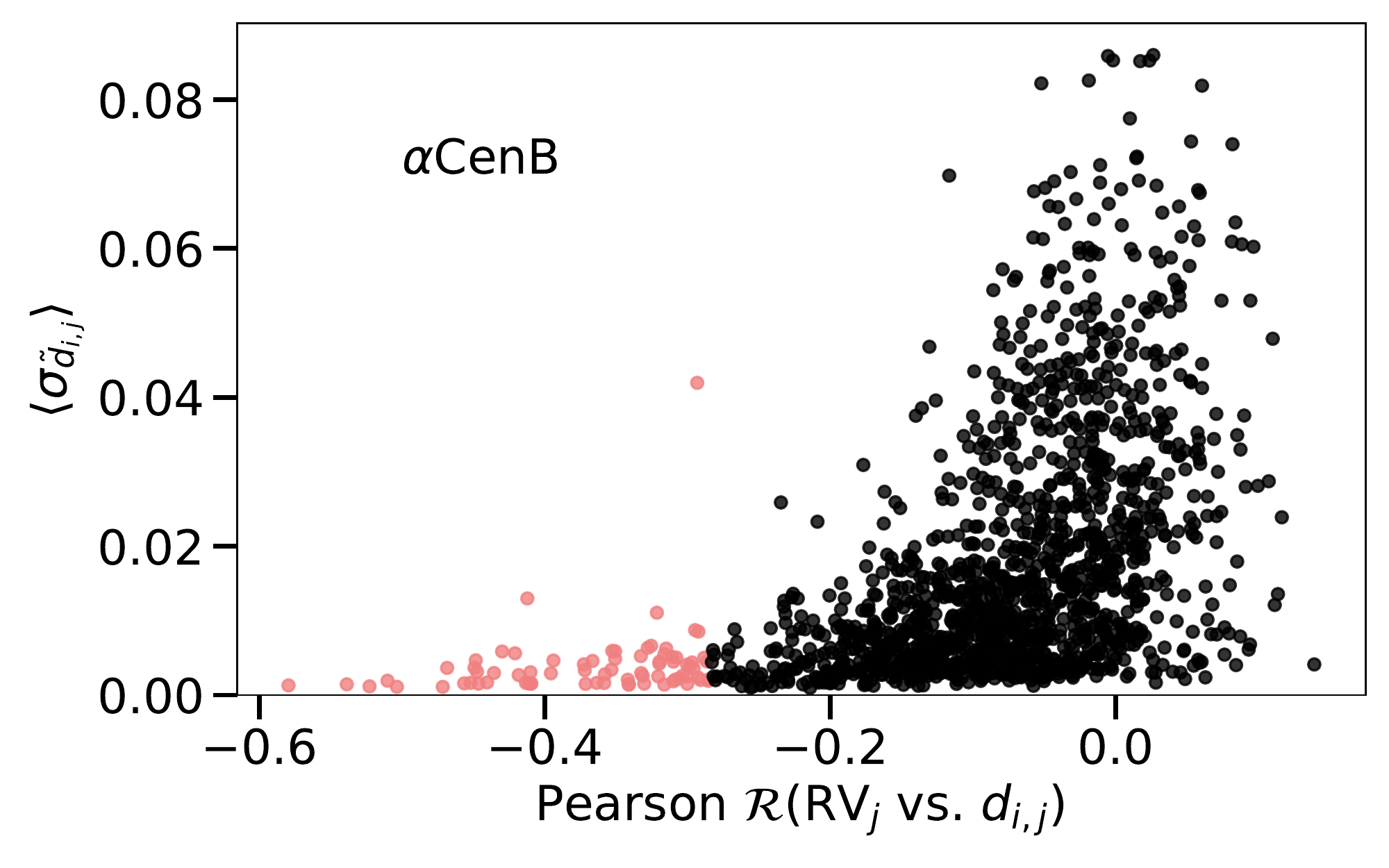}{0.5\textwidth}{} \fig{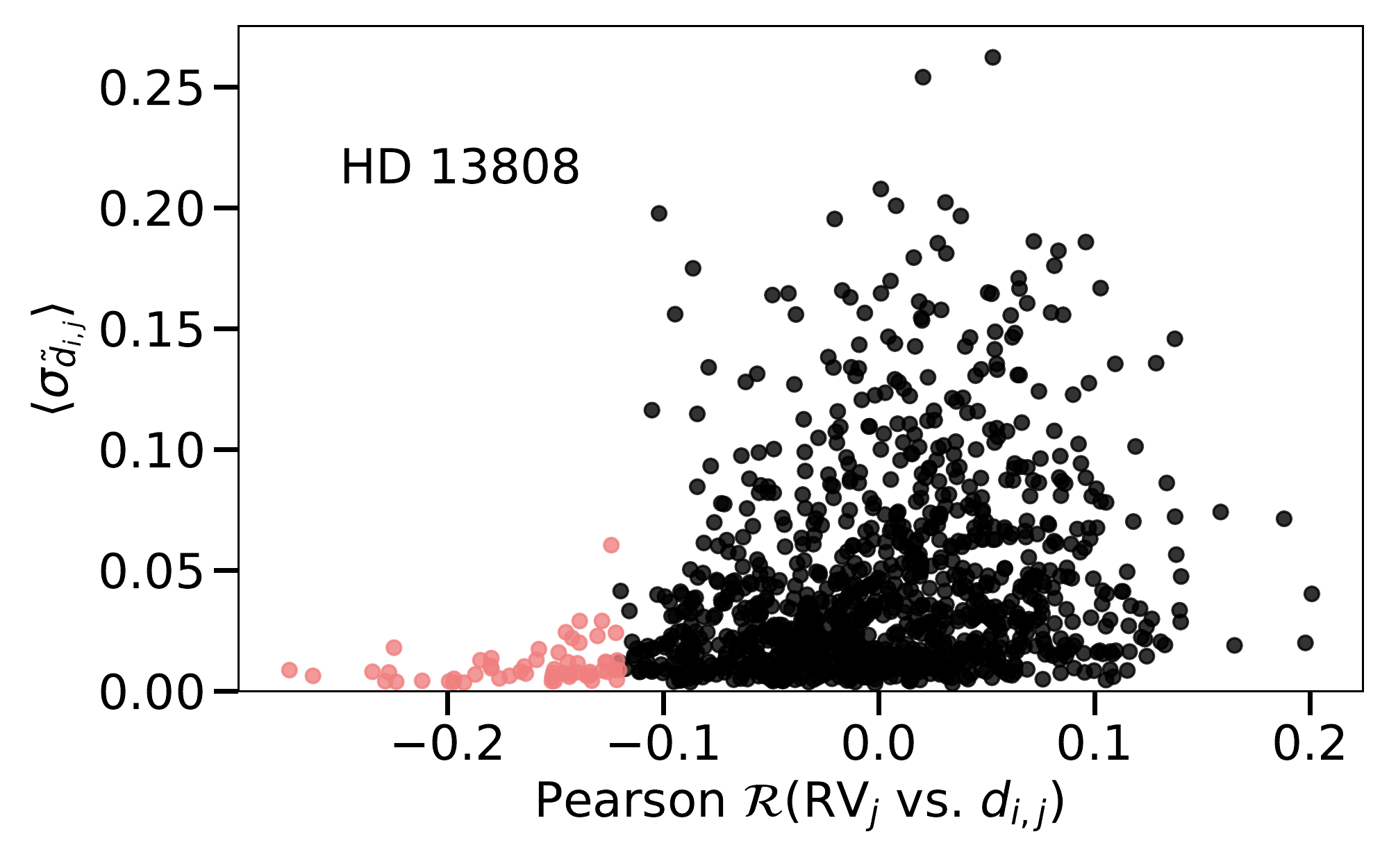}{0.5\textwidth}{} }

\caption{ Both $\alpha$CenB (\textit{left}) and HD~13808 (\textit{right}) possess clear subpopulations of activity-sensitive absorption lines (lines with $\mathcal{R}$(RV$_{j}$~vs.~$d_{i,j})<\rm5th$ percentile of each star's $\mathcal{R}$(RV$_{j}$~vs.~$d_{i,j}$) distribution are highlighted in red). For each star, we present the Pearson correlation coefficient $\mathcal{R}$ between RV$_j$ and line depth $d_{i,j}$ against each line's mean uncertainty in scaled depth.  }
\label{fig:depth_corr}
\end{figure*}

\begin{figure*}[]
\gridline{\fig{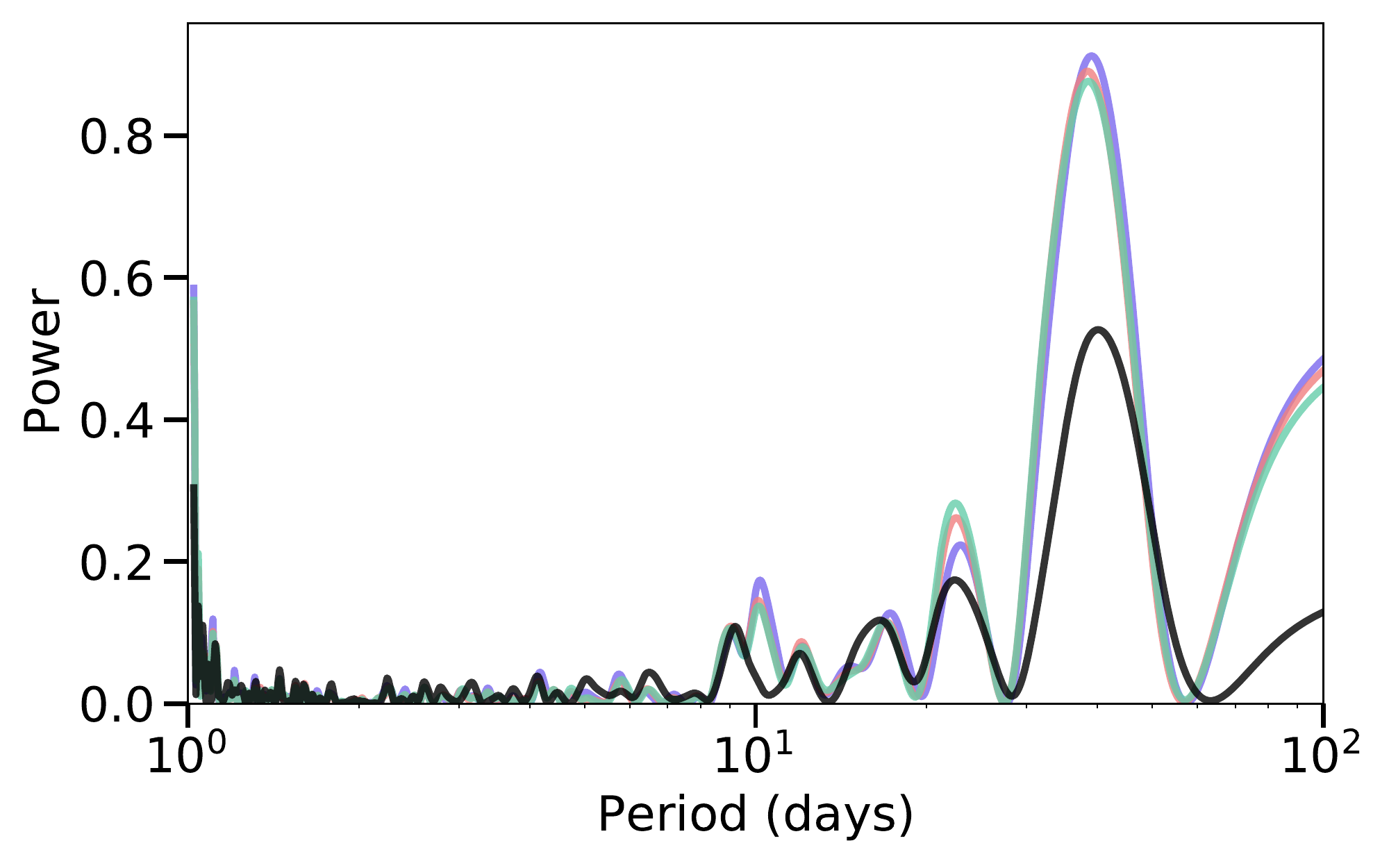}{0.5\textwidth}{} \fig{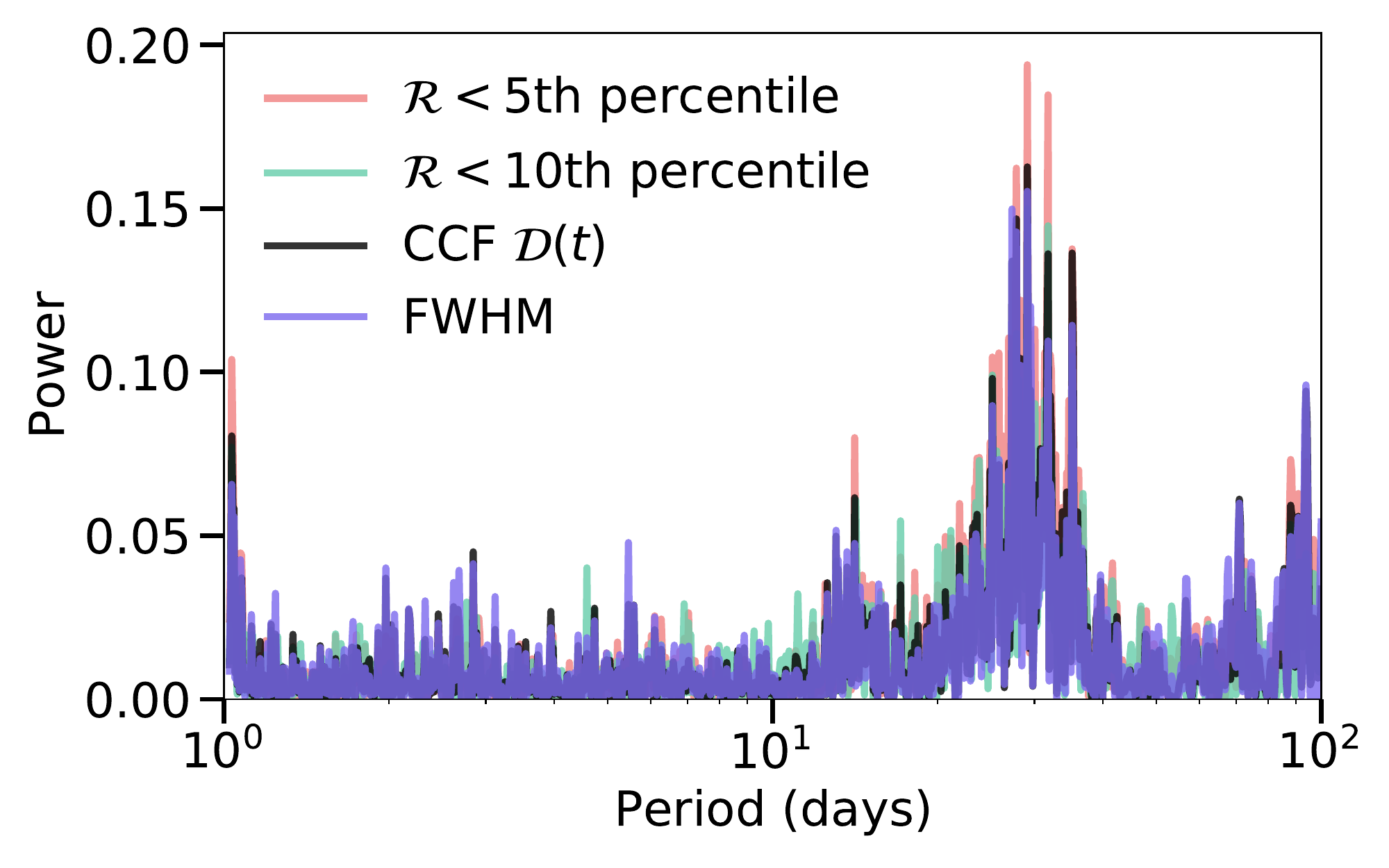}{0.5\textwidth}{} }

\caption{For both $\alpha$CenB and HD~13808, selecting a subpopulation of activity-sensitive absorption lines better isolates the star's stellar activity signal. We present Lomb-Scargle periodograms of the depth metric $\mathcal{D}$(t) for $\alpha$CenB (\textit{left}) and HD~13808 (\textit{right}) for two different spectral line subpopulations---$\mathcal{R}$(RV$_{j}$~vs.~$d_{i,j})<\rm 5th$ and $\rm10th$ percentile of each star's $\mathcal{R}$(RV$_{j}$~vs.~$d_{i,j}$) distribution---as well as each star's CCF-based depth metric CCF~$\mathcal{D}(t)$ and the FWHM of the CCF. } 
\label{fig:depth_selection}
\end{figure*}
For HD~13808, we considered all 11 yr of HARPS observations prior to the fiber change on the 1st of June 2015. After filtering for outliers in the HARPS DRS RVs (5$\sigma$ clipping for two iterations), filtering for continuum outliers, and filtering observations with an airmass greater than 1.5, we had 235/276 observations. 

\section{Line-by-line Analysis} \label{sec:methods}

\begin{figure*}[t]
\gridline{\fig{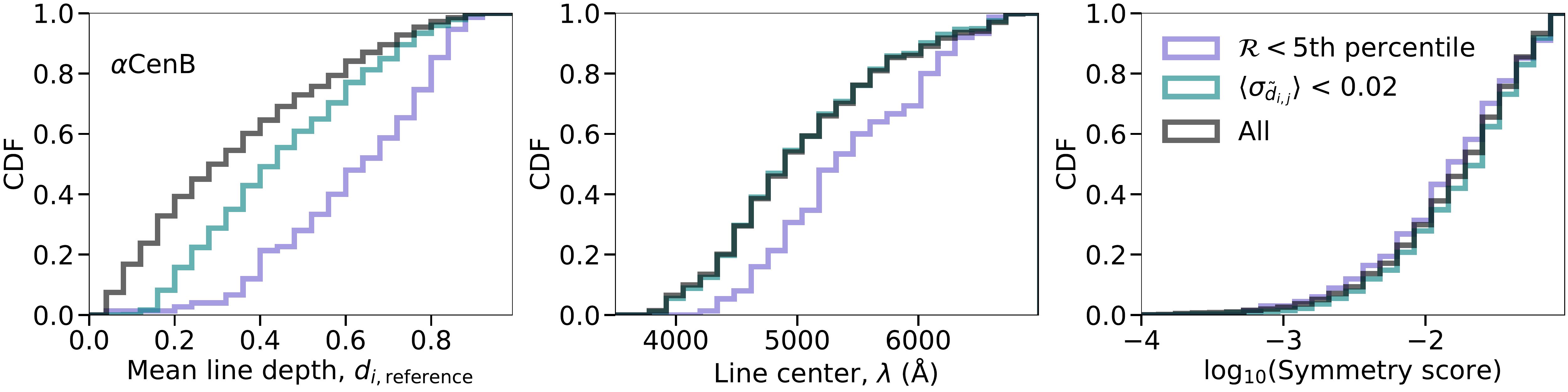}{\textwidth}{}}
\gridline{\fig{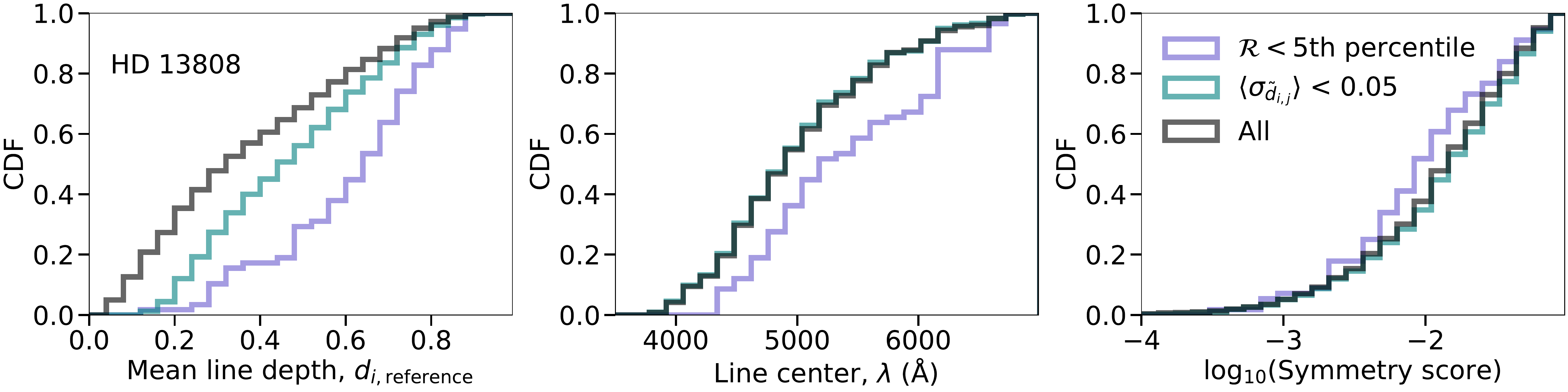}{\textwidth}{}}

\caption{The spectral lines shown to be sensitive to stellar activity tend to be deeper and favor longer wavelengths.  We present the cumulative distribution functions of reference spectrum line depth ($\tilde{d}_{i,\rm reference}$), central wavelength ($\lambda_i$), and symmetry for all lines and two subpopulations---activity-sensitive and low mean depth uncertainty---for $\alpha$CenB (\textit{top row}) and HD~13808 (\textit{bottom row}).}
\label{fig:depth_diag}
\end{figure*}

\begin{figure}[t]
\gridline{\fig{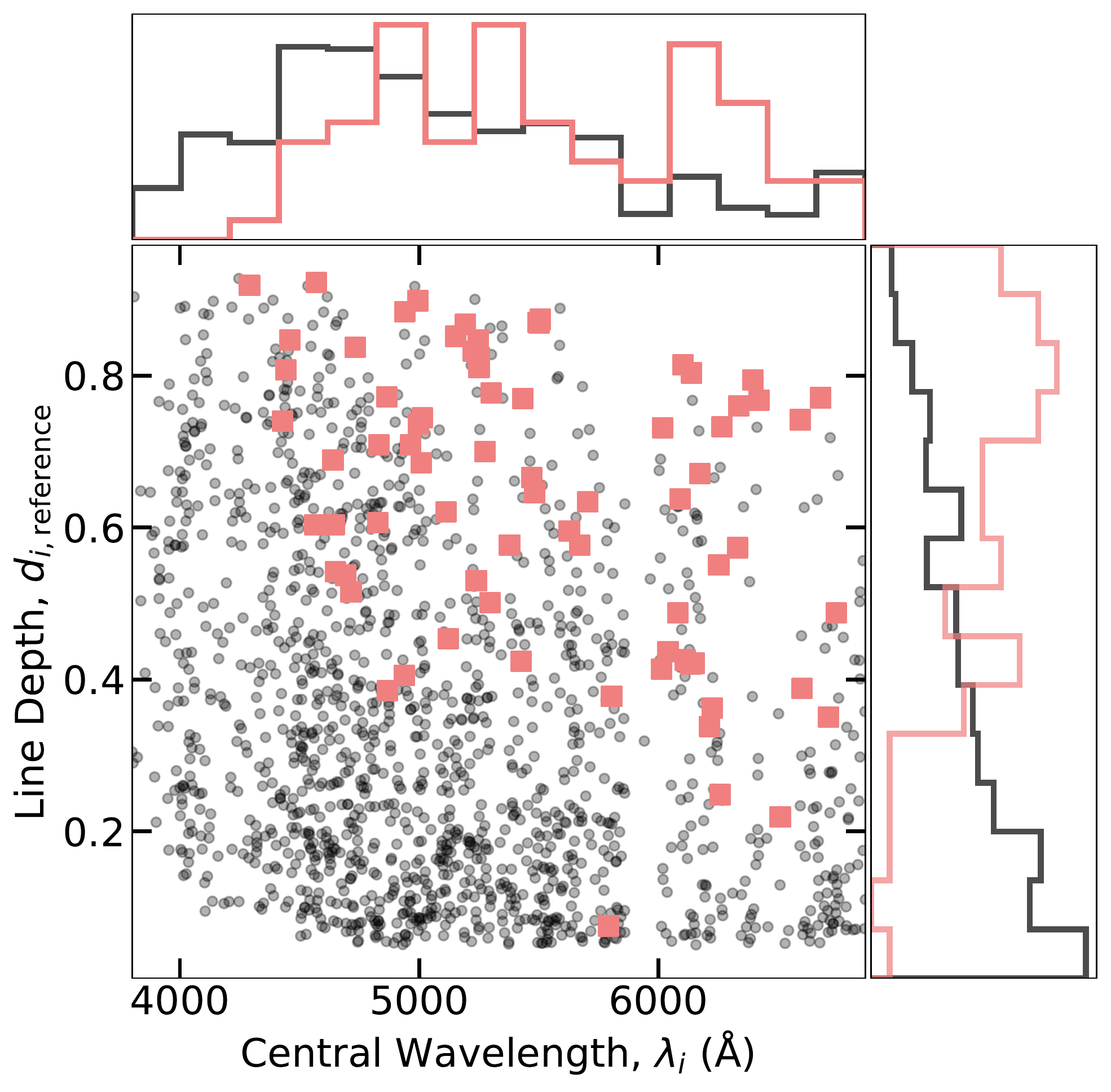}{\columnwidth}{} }

\caption{The subpopulation of activity-sensitive spectral lines in comparison with the entire population of spectral lines, for $\alpha$CenB. The activity-sensitive population is defined as all lines where $\mathcal{R}$(RV$_{j}$~vs.~$d_{i,j}$) is less than the $5$th~percentile of $\alpha$CenB's $\mathcal{R}$(RV$_{j}$~vs.~$d_{i,j}$) distribution. The activity-sensitive lines are shown as red squares, while all other lines are shown as black circles.}
\label{fig:depth_vs_wavelength}
\end{figure}

\begin{figure*}[ht]
\gridline{\fig{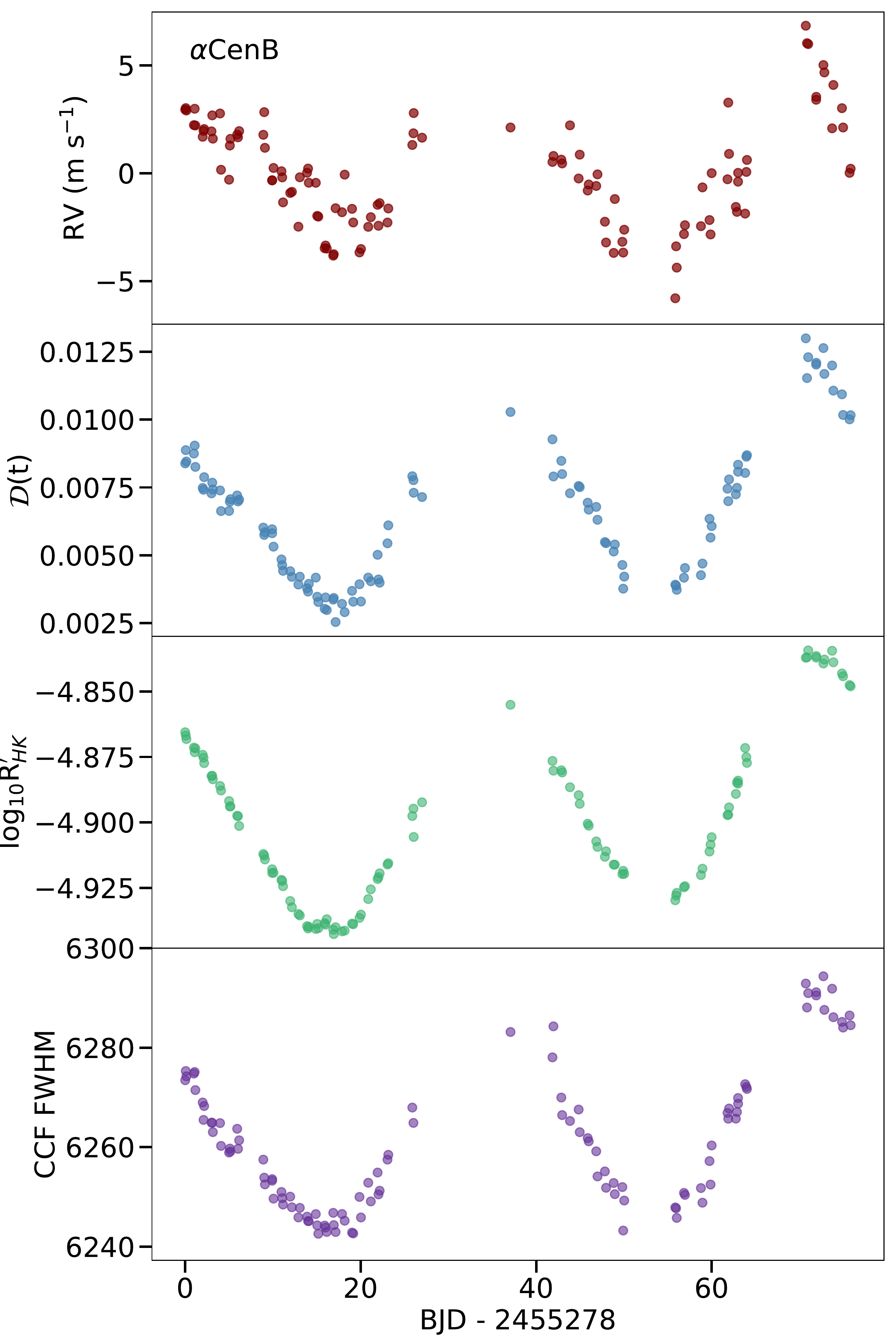}{.5\textwidth}{}\fig{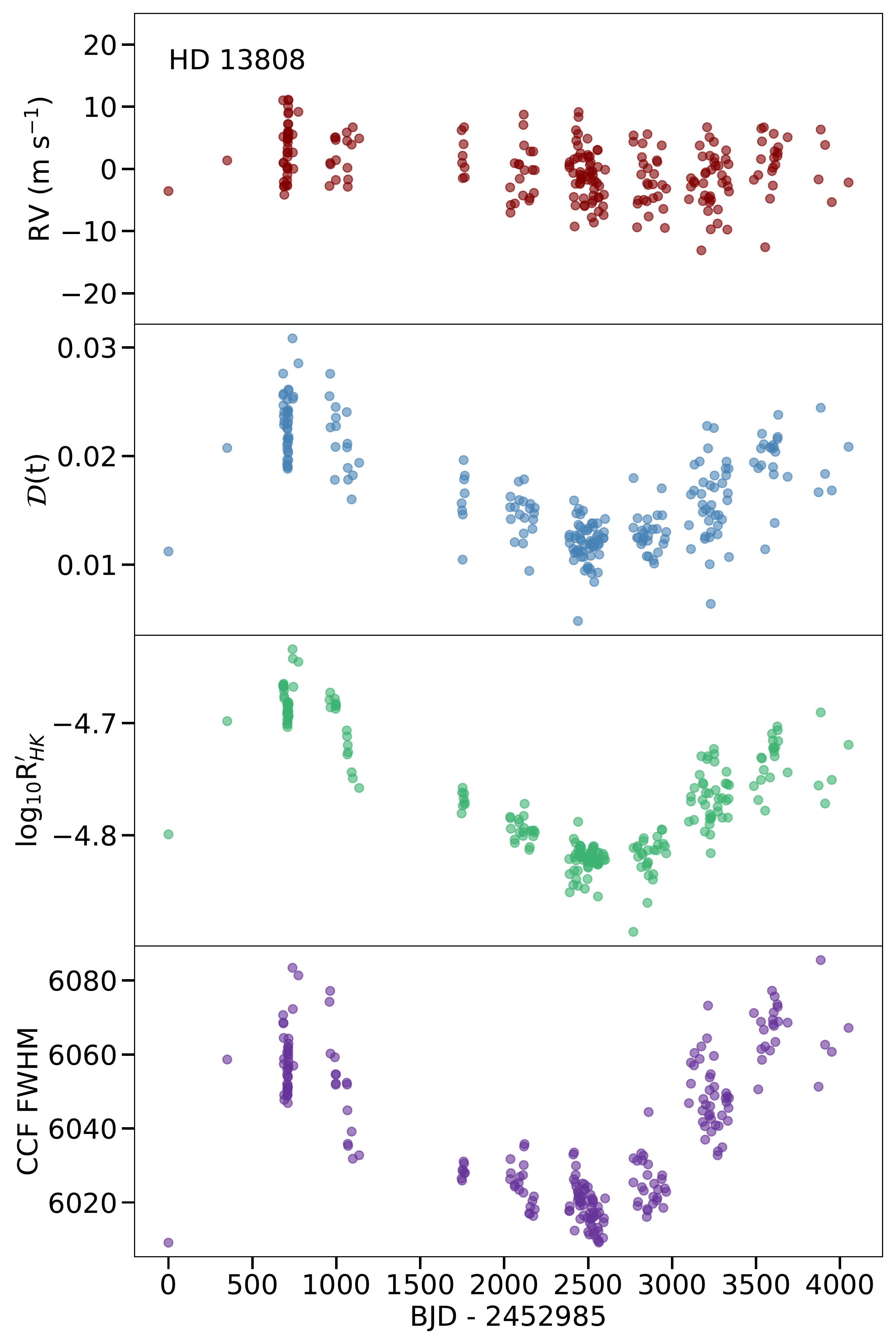}{.5\textwidth}{}}

\caption{For both $\alpha$CenB and HD~13808, the depth metric closely tracks the stellar activity signal. Here, we present time series of RV$_j$ compared with the depth metric $\mathcal{D}(t)$, log$R^{\prime}_{ HK}$, and FWHM of the CCF for $\alpha$CenB (\textit{left}) and HD~13808 (\textit{right}). Due to the high cadence of observations, the $\alpha$CenB data were binned to one tenth of a day.}
\label{fig:activity_comparison}
\end{figure*}

As developed by \cite{Dumusque2018}, the line-by-line method measures the RV of each spectral line independently using template matching. By taking the weighted average over all lines, the precise RV measurement for the entire spectrum can be obtained. For a given star, the line-by-line method can be broken down into three primary steps: first, construct a reference spectrum; second, generate a spectral line list from the reference spectrum or adopt a line list from a synthetic template; third, conduct template matching for each spectral line in each spectrum. Our implementation of the \cite{Dumusque2018} line-by-line method will be released for public use in Siegel J. et al. in preparation.

We generated the reference spectrum for a given star by co-adding the systematics-corrected spectra on a common wavelength solution; the common wavelength solution is taken to be that of the spectrum with the lowest barycentric Earth RV, and linear interpolation was used to evaluate all spectra on this common solution.

Next, we generated a catalog of absorption lines following the methods detailed in Appendix A of \cite{Cretignier2020}. Using a given star's reference spectrum, absorption lines were identified in terms of local minima and local maxima. For two neighboring local maxima at $\lambda_{i, \rm left}$ and $\lambda_{i, \rm right}$, with a local minima between them at $\lambda_{i}$, we treated $\lambda_{i}$ as the line's center and adopted a window width of $\min(\lambda_i-\lambda_{i, \rm left},\lambda_{i,\rm right}-\lambda_i)$. We then rejected all lines where $\lambda_{i, \rm left}$ and $\lambda_{i, \rm right}$ were within 10 pixels of each other and demanded that the line's depth (defined below in Eqn. \ref{equ:depth}) be greater than 0.05. For $\alpha$CenB and HD~13808, we identified a total of $\sim$7,500 and $\sim$6,500 lines, respectively, which reduced to $\sim$5,500 and $\sim$5,000 when overlap of spectral orders was taken into account. For lines found in multiple spectral orders, we considered the spectral order for which the line had the lowest mean RV uncertainty, $\langle \sigma_{ \text{RV}_{i,j} } \rangle$. Given the similar spectral types of $\alpha$CenB and HD~13808, we attribute the difference in the number of detected lines for each star to the higher average signal-to-noise ratio of $\alpha$CenB's spectra; for order 50, $\alpha$CenB has an average signal-to-noise ratio of 320, while HD~13808 has an average of 110. Consistent with this explanation, we found more shallow spectral lines for $\alpha$CenB than HD~13808; for $\alpha$CenB, we identified $\sim 200$ spectral lines with $d_{i,\rm reference}<0.1$, while for HD~13808, we identified only $\sim 100$ such spectral lines.

The radial velocity RV$_{i,j}$ was measured by numerically solving Eqn. 2 of \cite{Bouchy2001}. For the $i$th line of the $j$th spectrum,
\begin{equation}
    S_{i,j}(\lambda) = A \Big[ S_{i,\rm reference}(\lambda) + \frac{\partial S_{i,\rm reference}(\lambda)}{\partial \lambda} \delta \lambda \Big],
\end{equation}
where $S_{i,j}$ are the continuum normalized pixel flux values within the line window, and $S_{i,\rm reference}$ are the reference spectrum flux values over the same interval. The reference spectrum was evaluated on the wavelength solution of the $j$th spectrum via cubic spline interpolation. RV$_{i,j}$ was then given by
\begin{equation}
    \text{RV}_{i,j}=\frac{c}{\lambda}\delta\lambda.
\end{equation}
Uncertainty in RV$_{i,j}$ was inferred by propagating the uncertainties in $A\delta\lambda$ and $A$ from the least-squares fit,
\begin{equation}
    \sigma_{ \text{RV}_{i,j} } = \frac{c}{\lambda}\delta\lambda \sqrt{ \Big( \frac{\sigma_{A\delta\lambda}}{A\delta\lambda} \Big)^2 + \Big( \frac{\sigma_A}{A} \Big)^2 }.
\end{equation}

We also extracted the depth of each spectral line, defined as
\begin{equation}
    \label{equ:depth}
    d_{i,j} = 1 - \frac{\min(S_{i,j})}{C_{i,j}},
\end{equation}
where $C_{i,j}$ is the continuum flux, defined as the average of the leftmost and rightmost pixel fluxes in the line's window; Eqn.~\ref{equ:depth} describes the depth of a spectral line relative to the local continuum, which is equivalent to the depth relative to the continuum spectrum for isolated lines. For high-signal-to-noise lines, greater precision may be achieved by inferring the minimum flux by cubic spline interpolation; however, we found native pixel sampling yields results nearly-indistinguishable from those of cubic spline interpolation and avoids potential instabilities for low-signal lines. Lastly, we defined the normalized depth parameter, 
\begin{equation}
    \tilde{d}_{i,j} =  d_{i,j} / d_{i,\rm reference} 
\end{equation}
where $d_{i,\rm reference}$ is the line's depth in the reference spectrum (effectively the line's mean depth). Uncertainty in the depth parameter was calculated via propagation of the flux uncertainties. 

Following our initial line-by-line analysis, we proceeded to clean the dataset of outlier measurements. Similar to the sigma-clipping procedures of \cite{Dumusque2018}, for each observation we performed a 4$\sigma$ clipping procedure (with two iterations) on RV$_{i,j}$, $\sigma_{ \text{RV}_{i,j}}$, $\chi_{i,j}^2$ of the $A\delta\lambda$ and $A$ fit, $\tilde{d}_{i,j}$, and $\sigma_{\tilde{d}_{i,j}}$. We then rejected all lines for which more than 1\% of the measurements were rejected by the above filters, and we removed all lines within 48 pixels of a telluric line. Telluric lines were identified using a sample \texttt{TAPAS} model spectrum \citep{Bertaux2014} for Keck Observatory (30$^{\circ}$ zenith). Spectral regions in the vicinity of TAPAS features with a normalized contamination depth of 1\% or greater were rejected; we found that adopting a 0.1\% depth threshold or filtering all lines in our line list with a central wavelength $>$5,000$~ \rm \AA$ did not yield RVs significantly different from our baseline 1\% threshold. Following Table C.1 of \cite{Cretignier2020}, we also rejected asymmetric absorption lines, because such lines have been shown to be anticorrelated with stellar activity. After applying these filters, we retained $\sim$1,500 and $\sim$1,100 lines for $\alpha$CenB and HD~13808, respectively. Greater than $400$ lines appear in both star's line lists. 

Using the cleaned dataset, we next combined the line-by-line information into bulk properties. For each observation, we conducted a weighted average to derive the cumulative RV:
\begin{equation}
        \label{equ:weighted_average_rv}
       \text{RV}_j = \sum_i \frac{\text{RV}_{i,j}}{\sigma^2_{\text{RV}_{i,j}}} /  \sum_i \frac{1}{\sigma^2_{\text{RV}_{i,j}}} .
\end{equation}

\subsection{The depth metric}
\label{sec:depth}

\cite{Wise2018} demonstrated that, for a subpopulation of visually identified spectral lines, there is a strong correlation between line depth and stellar activity as traced by the Ca II H\&K index. Here, we conducted a data-driven search for such spectral lines. For a given stellar target, we calculated the Pearson correlation coefficient $\mathcal{R}$ between the bulk RV, RV$_j$, and line depth $d_{i,j}$. As seen in Figure \ref{fig:depth_corr}, both $\alpha$CenB and HD~13808 possess substantial subpopulations of activity-sensitive lines. We used RV$_j$ to calculate each line's activity sensitivity, because standard activity indices (e.g.,  log$R^{\prime}_{HK}$ and FWHM of the CCF) have been shown to lag in time relative to the RV signal \citep{Cameron2019}. However, planetary signals dilute the correlation between depth and the activity contribution to measured RVs, potentially limiting the scope of this approach; we discuss methods of selecting activity-sensitive lines for stars with strong planetary signals in Section \ref{sec:concl}. 

To reflect our uncertainty in $d_{i,j}$ and RV$_j$, we conducted 1,000 Monte Carlo trials for each line. For a given trial, we drew an independent sample of depth and RV values, assuming Gaussian uncertainties. A given line's Pearson correlation coefficient $\mathcal{R}$ was taken as the average over all 1,000 trials.

For $\alpha$CenB and  HD~13808, the subpopulation of activity-sensitive lines was defined as all lines satisfying $\mathcal{R}<5$th~percentile of each star's $\mathcal{R}$ distribution, i.e. only the most activity-sensitive lines were used. As shown in Figure \ref{fig:depth_selection}, these adopted thresholds maximize each star's stellar activity signal, while still maintaining a significant population of spectral lines; $>60$ lines for $\alpha$CenB and $>50$ lines for HD~13808. Although line depth variations are reflected in each observation's computed CCF, which includes thousands of lines, we clearly see in Figure \ref{fig:depth_selection} that the signal is greatly amplified by selecting an activity-sensitive subpopulation. The CCF-based depth metric (CCF~$\mathcal{D}(t)$) was obtained by fitting each HARPS DRS CCF to a Gaussian model with a constant continuum and feeding the resulting depth time series into Eqn. \ref{equ:Dt} (described below); inferring the depth of the CCF with cubic spline interpolation, instead of the Gaussian model, yielded negligible differences in CCF~$\mathcal{D}(t)$ for both $\alpha$CenB and HD~13808.

In Figure \ref{fig:depth_diag}, we present the distributions of reference spectrum line depth $d_{i, \rm reference}$, central wavelength $\lambda_i$, and symmetry score for the activity-sensitive lines alongside the general population. For a given line, the symmetry score is defined as the absolute difference between the leftmost and the rightmost pixel fluxes in the line's window, divided by the mean of the leftmost and the rightmost pixel fluxes; a highly symmetric line has a symmetry score near zero. For both $\alpha$CenB and HD~13808, the activity-sensitive population is comprised mostly of deeper lines at longer wavelengths, with a slight preference for symmetric lines. Since $\mathcal{R}$ is highly dependent on $\langle \sigma_{\tilde{d}_{i,j}} \rangle$ (see Figure \ref{fig:depth_corr}), we also isolated the subpopulation of lines with low $\langle \sigma_{\tilde{d}_{i,j}} \rangle$ for each star. Although the preference for deep lines is consistent with the preference for small uncertainties in depth, the low $\langle \sigma_{\tilde{d}_{i,j}} \rangle$ populations do not significantly differ from the general population with regard to central wavelength or symmetry. This preference for longer wavelengths may reflect the higher density of lines at shorter wavelengths (see Figure \ref{fig:depth_vs_wavelength}); line blending makes it difficult to measure a given line's activity signal.

Since a single spectral line possesses only a small amount of information, we combined the activity-sensitive lines into the depth metric. Adapting the formalism of \cite{Aigrain2012}, for a given star we calculated each spectrum's weighted average line depth over our selected subpopulation and rescaled the resulting time series to introduce the depth metric $\mathcal{D}(t)$:
\begin{align}
    \tilde{d}(t) &= \sum_i \frac{\tilde{d}_{i,j}}{\sigma^2_{\tilde{d}_{i,j}}} /  \sum_i \frac{1}{\sigma^2_{\tilde{d}_{i,j}}},\\
    \label{equ:d_0}
    \tilde{d}_0 &= \tilde{d}_{\rm max} + \sigma_{\tilde{d}},\\
     \label{equ:Dt}
    \mathcal{D}(t) &= 1 - \frac{\tilde{d}(t)}{\tilde{d}_0},
\end{align}
where $\tilde{d}_{\rm max}$ is the maximum depth metric of the $ \tilde{d}(t)$ time series and $\sigma_{\tilde{d}}$ is the standard deviation of the time series. As seen in Figure \ref{fig:activity_comparison}, the depth metric captures the stellar activity cycles of $\alpha$CenB and HD~13808. For comparison, we also present the log$R^{\prime}_{HK}$ and CCF FWHM activity indices for each star; log$R^{\prime}_{HK}$ was obtained by first calculating the $\mathcal{S}_{\rm index}$ via \cite{GomesdaSilva2018} and rescaling the $\mathcal{S}_{\rm index}$ by stellar color following \cite{Noyes1984}.

Compared to standard activity indices, $\mathcal{D}(t)$ has several advantages. Unlike log$R^{\prime}_{HK}$ and H$\alpha$, the depth metric covers a wide range of wavelength space and should be accessible to spectrometers spanning different wavelength regions. By likely probing temperature changes on the stellar surface through Zeeman-splitting \citep{Wise2018}, the depth metric more closely tracks the stellar surface than chromospheric indices. Unlike diagnostics of the CCF, namely FWHM and BIS, the depth metric only considers lines shown to be activity-sensitive, more clearly isolating the stellar activity signal (see Table \ref{tab:correlation_comp}). Table \ref{tab:correlation_comp} reports the Pearson correlation coefficient between the observed RVs and each activity index for $\alpha$CenB and HD~13808; of the activity indices we considered, the depth metric has the strongest correlation with the measured RVs, for both stars.

However, linear correlation is a useful, but imperfect, standard for evaluating the quality of an activity index. The diagnostic power of an activity index should be weighted by considering both its application to RV detrending models and its accessibility (e.g., photometry facilitates high-quality activity tracking but is unavailable for the majority of RV measurements). To determine the quality of the depth metric, we conducted a comparative study of several activity indices (FWHM of the CCF, log$R^{\prime}_{HK}$, and the depth metric) for a variety of detrending models, see Section~\ref{sec:model_comparison}. In Section~\ref{sec:concl}, we discuss the accessibility of the depth metric and outline potential use cases.

\begin{deluxetable*}{cccccc}[t]
\tablecaption{ Activity index correlation comparison \label{tab:correlation_comp}}
\tablehead{
 \colhead{} &  \colhead{} & \colhead{RV} & \colhead{log$_{10}$R$^{\prime}_{HK}$} & \colhead{FWHM} & \colhead{$\mathcal{D}(t)$}
}
\startdata
$\alpha$CenB & & & & & \\
 & RV & --- & 0.776 & 0.770 & 0.813\\
 & log$_{10}$R$^{\prime}_{HK}$ & 0.776 & --- & 0.974 & 0.979\\
 & FWHM & 0.770 & 0.974 & --- & 0.981\\
 & $\mathcal{D}(t)$ & 0.813 & 0.979 & 0.981 & ---\\
\hline
HD~13808 & & & & & \\
 & RV & --- & 0.369 & 0.310 & 0.446\\
 & log$_{10}$R$^{\prime}_{HK}$ & 0.369 & --- & 0.835 & 0.836\\
 & FWHM & 0.310 & 0.835 & --- & 0.681\\
 & $\mathcal{D}(t)$ & 0.446 & 0.836 & 0.681 & ---\\
\enddata
\tablecomments{For each star, we report the Pearson correlation coefficient $\mathcal{R}$ between each activity index, as well as the correlation coefficient between each activity index and the radial velocity measurements. }

\end{deluxetable*}

\section{Stellar Activity Mitigation Methods}
\label{sec:mitigation}

As instruments approach the precision level necessary to detect an Earth twin, stellar activity can still generate m~s$^{-1}$ level noise and induce planet-like RV signatures. The line-by-line technique has the potential to greatly reduce such noise, since each spectral line responds differently to stellar activity depending on the details of the line transition. Here, we describe four stellar activity corrections: line selection \citep{Dumusque2018}, formation depth \citep{Cretignier2020}, FF$^{\prime}$ \citep{Aigrain2012}, and Gaussian progress (GP) regression \citep{Haywood2014}. In Section \ref{sec:model_comparison}, we conduct a thorough comparison of these mitigation methods using $\alpha$CenB and HD~13808 HARPS observations. 

\subsection{Line Selection Method}
\label{sec:line_selection}

\cite{Dumusque2018} first demonstrated that, by selecting a population of activity-insensitive lines, the RV scatter can be reduced; in the case of $\alpha$CenB, \cite{Dumusque2018} achieved a factor of $\sim 1.5$ scatter reduction. For this initial investigation, the subpopulation of activity-insensitive lines was selected based on each line's mean uncertainty in RV$_{i,j}$ and each line's Pearson correlation coefficient between RV$_{i,j}$ and RV$_{j}$.

We selected our activity-insensitive set of lines through a similar filter. For a given star, we calculated each line's mean correlation coefficient between RV$_{i,j}$ and RV$_{j}$ from 1,000 Monte Carlo trials. We then defined the activity-insensitive population as all lines satisfying $\mid \mathcal{R} \mid<0.1$ and $\langle \sigma_{ \text{RV}_{i,j}} \rangle<50$~m s$^{-1}$. Activity-corrected RVs were determined by recalculating RV$_j$ using only these activity-insensitive lines.

\subsection{Formation Depth Method}
\label{sec:formation_depth}
Utilizing the relationship between a line's formation depth and the magnitude of convective blueshift, \cite{Cretignier2020} introduced a stellar activity correction based on formation depth. For a population of activity-sensitive lines, \cite{Cretignier2020} first demonstrated that the magnitude of a given line's sensitivity to activity is inversely proportional to that line's depth in the reference spectrum (effectively the line's mean depth). A subpopulation of shallow activity-sensitive lines thus show a greater RV scatter due to stellar activity than a subpopulation of deep activity-sensitive lines. As a result, the difference between the weighted average RV of the shallow subpopulation (RV$_{j,\rm shallow}$) and the deep subpopulation (RV$_{j,\rm deep}$) acts as an activity proxy. The stellar activity corrected RVs are then RV$_j-$(RV$_{j,\rm shallow}-$RV$_{j,\rm deep}$).

Similar to the line selection method, there is considerable flexibility, and likely great room for improvement, in defining these subpopulations of lines. For $\alpha$CenB, we defined the population of active lines as those where the correlation coefficient between RV$_j$ and RV$_{i,j}$ is greater than 0.2, and for HD~13808 we adopted a threshold of 0.1. From this population of active lines, we defined the shallow population as all lines satisfying $0.1<d_{i,\rm reference}<0.4$ and the deep population as those satisfying $d_{i,\rm reference}>0.6$; these depth thresholds were applied to both $\alpha$CenB and HD~13808.

\subsection{FF$^{\prime}$ Method}

The FF$^{\prime}$ model, first described by \cite{Aigrain2012}, uses photometric flux and a spot model to predict the RV signal due to stellar activity. For a normalized flux time series $F(t)$, the FF$^{\prime}$ model predicts the effects of both convective blueshift $\Delta \text{RV}_c$ and the rotation of a single spot $\Delta \text{RV}_r$, 
\begin{align}
    \Delta \text{RV}_c &=   F^2(t)\delta V\kappa/f,\\
    \Delta \text{RV}_r &= -F(t)\dot{F}(t)R_{\star}/f,
\end{align}
where $R_{\star}$ is the stellar radius, $f$ is the relative flux drop for a spot at the disk center, $\delta V$ is the difference between the convective blueshift in the unspotted photosphere and that within the magnetized area, and $\kappa$ is the ratio of the magnetized area to the area of the spot. The combined prediction for the stellar RV signal is
\begin{equation}
    \text{FF}^{\prime}(t) = \Delta \text{RV}_c + \Delta \text{RV}_r.
\end{equation}

\cite{Giguere2016} introduced the HH$^{\prime}$ method, which uses the H$\alpha$ index as a photometric proxy. Using the same model as \cite{Aigrain2012}, with the addition of two free parameters  to linearly scale H$\alpha$ to flux and a smoothing term to compensate for low cadence, \cite{Giguere2016} found the HH$^{\prime}$ method outperformed linear decorrelation but yielded a factor of two more scatter than the FF$^{\prime}$ model. \cite{Giguere2016} postulated that the inclusion of more activity indicators could improve the HH$^{\prime}$ method. 

Using a modified FF$^{\prime}$ framework, we applied the methods of \cite{Aigrain2012} to the depth metric $\mathcal{D}(t)$, the FWHM of the CCF, and log$R^{\prime}_{HK}$. Treating the relative amplitudes of the spot rotation and convective blueshift effects as a free parameter and allowing for a linear relationship between $\mathcal{A}(t)$ and flux, our modified FF$^{\prime}$ RV prediction is
\begin{align}
    \label{equ:DD}
    \text{FF}^{\prime}(t) = &-\alpha(\mathcal{A}(t) +\beta)\dot{\mathcal{A}}(t)/f  \nonumber \\ 
    &+ \gamma (\mathcal{A}(t)+\beta)^2 /f \nonumber \\ 
    &+ \mathcal{C}_2 t + \mathcal{C}_1,
\end{align}
where $\mathcal{A}(t)$ is a normalized activity index time series (defined analogously to Eqn. \ref{equ:Dt}), $f$ is included to conveniently scale $\alpha$ and $\gamma$, $\beta$ is the zero-point of the assumed linear relationship between  $\mathcal{A}(t)$ and photometric flux, and $C_1$ and $C_2$ are an arbitrary zero-point and linear drift, respectively; following \cite{Aigrain2012}, we assumed $f$ is given by
\begin{equation}
    f=\frac{\mathcal{A}_0-\min ( \mathcal{A}(t))}{\mathcal{A}_0},
\end{equation}
where $\mathcal{A}_0$ is defined analogously to Eqn. \ref{equ:d_0}.

To estimate $\dot{\mathcal{A}}(t)$ for low-cadence targets, we smoothed $\mathcal{A}(t)$ using a smoothing parameter $\sigma_t$. For the $j$th observation, the smoothed $\mathcal{A}_j$ is the weighted average of the entire activity metric time series, where the weights are assigned via a Gaussian centered at $t_j$ (the time of the $j$th observation) with a standard deviation of $\sigma_t$. The smoothed activity metric is fit with a cubic spline and $\dot{\mathcal{A}}(t)$ is determined analytically. Our modified FF$^{\prime}$ model has six free parameters: $\alpha$, $\beta$, $\gamma$, $\mathcal{C}_1$, $\mathcal{C}_2$, and $\sigma_t$. 

\begin{figure*}[t]
\gridline{\fig{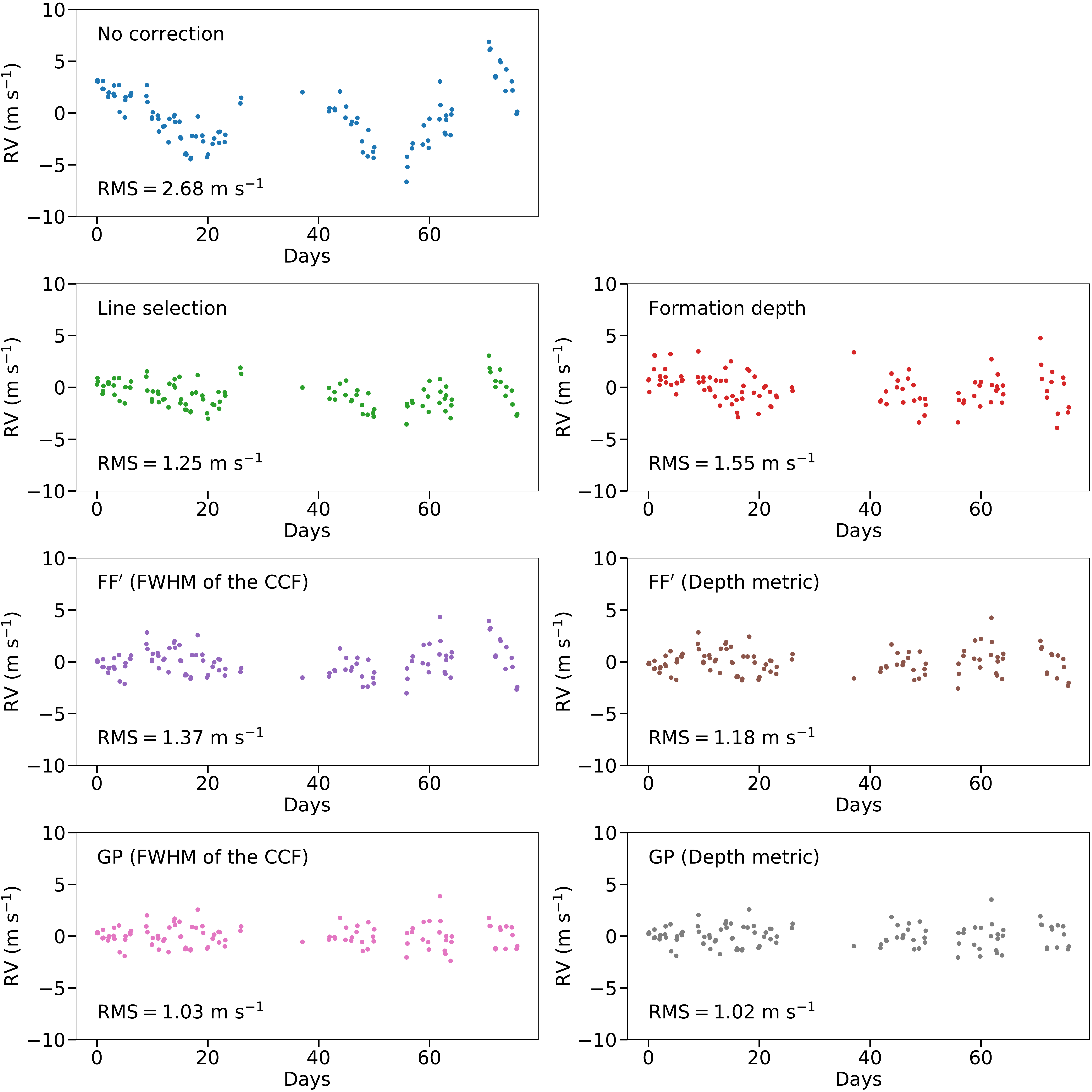}{.9\textwidth}{}}
\caption{Each method successfully reduces the amplitude of $\alpha$CenB's stellar activity signal, with the depth metric based methods performing at least as well as the FWHM of the CCF-based methods. We present the RVs of the 2010 HARPS $\alpha$CenB observations under six different stellar activity mitigation methods, as well as the uncorrected case. The mitigation methods are described in Section \ref{sec:mitigation}. }
\label{fig:alphaCenB_mitigation}
\end{figure*}

\begin{figure*}[t]
\gridline{\fig{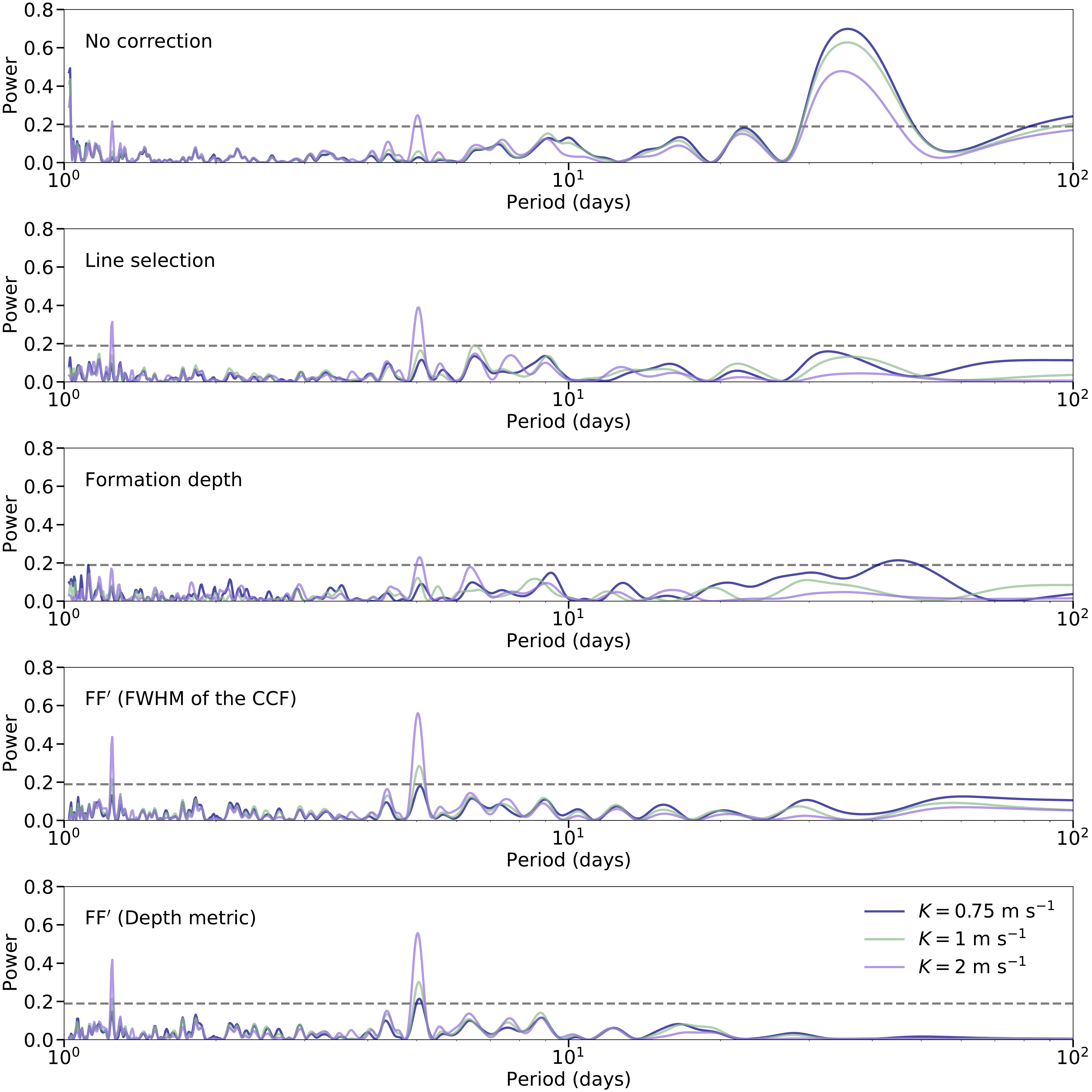}{.9\textwidth}{}}
\caption{In terms of signal recovery, the FF$^{\prime}$ models outperform the line selection and formation depth methods. We present the Lomb-Scargle periodograms for the HARPS $\alpha$CenB observations under four stellar activity mitigation methods, as well as the uncorrected case; to mimic a blind planet search, we omit the Keplerian component in the RV models. Each color corresponds to a different RV semi-amplitude of the injected five-day planet. The horizontal dashed line demarcates the 1\% false-alarm-probability level. RVs were binned to one tenth of a day.}
\label{fig:alphaCenB_injection}
\end{figure*}

\begin{figure*}[t]
\gridline{\fig{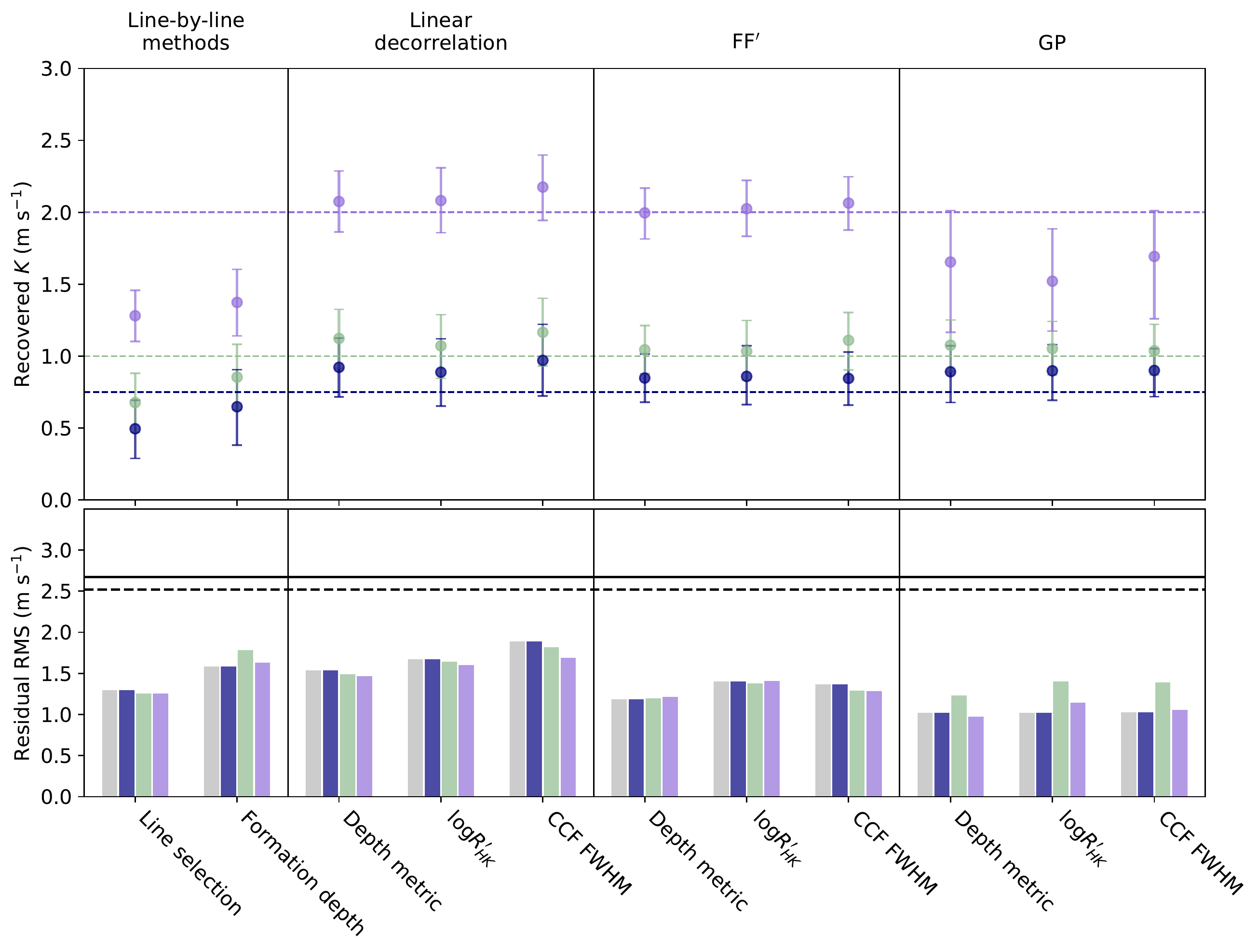}{.95\textwidth}{}}
\caption{For a range of activity mitigation models, the depth metric performs at least as well as FWHM of the CCF and log$R^{\prime}_{HK}$. We present the recovered RV semi-amplitude $K$ (m~s$^{-1}$) and residual RMS  (m~s$^{-1}$) for each combination of mitigation method, activity index, and injection scenario. \textit{Top panel}: the recovered RV semi-amplitude for the $K=0.75,1$, and $2$~m~s$^{-1}$ injection cases are shown in blue, green, and purple, respectively. The injected RV semi-amplitudes are shown as dashed horizontal lines. \textit{Bottom panel}: the residual RMS for each injection case. The RMS for the null injection case is shown in gray. The RMS of the uncorrected line-by-line RVs and the DRS RVs are shown as solid and dashed horizontal lines, respectively.}
\label{fig:injection_recovery}
\end{figure*}

\subsection{Gaussian process regression}
\label{sec:gp}
Gaussian processes (GPs) model a stochastic process using a parametric form for the covariance matrix. Prior studies have modeled structured stellar noise in RVs using a quasi-periodic kernel \citep{Haywood2014, Dai2017, Ahrer2021}. Following \cite{Haywood2014}, we define a given element of the covariance matrix as
\begin{align}
    C_{n,m} &= \eta_1^2 \exp \Bigg[ - \frac{(t_n-t_m)^2 }{\eta_2^2} - \frac{\sin^2 (\pi \mid t_n - t_m\mid/\eta_3)}{2 \eta^2_4}\Bigg] \nonumber\\
    &+\delta_{n,m}\Big[ \sigma_{\text{RV}_n}^2 + \sigma_{\text{jitter}}^2\Big],
\end{align}
where $n$ and $m$ index over the observations, $\eta_1$ is the activity amplitude, $\eta_2$ is the correlation length scale, $\eta_3$ is the period of the covariance, $\eta_4$ is the relative significance of the squared exponential and periodic components, $\delta_{n,m}$ is the Kronecker delta function, and $\sigma_{\text{jitter}}$ is a white-noise jitter term.

The logarithmic likelihood function is
\begin{equation}
    \ln \mathcal{L} = - \frac{N}{2} \log 2 \pi - \frac{1}{2} \log \mid \boldsymbol{C} \mid - \frac{1}{2} \boldsymbol{r}^\text{T} \boldsymbol{C}^{-1} \boldsymbol{r},
\end{equation}
where $N$ is the number of observations, $\boldsymbol{C}$ is the covariance matrix, and $\boldsymbol{r}$ is the residual between the observed RVs and the model RVs.

For a given star and activity index, we modeled the structured stellar noise with a GP model conditioned on the activity index. We first modeled the activity index with the quasi-periodic GP kernel. For all activity indices, we adopted Jeffreys priors on $\eta_1$, $\eta_2$, and $\sigma_{\text{jitter}}$. For $\eta_4$, we adopted a Gaussian prior of $0.5 \pm 0.05$ \citep{Kosiarek2020}, and for $\eta_3$ we adopted a Gaussian prior centered on the star's rotation period. Through Markov Chain Monte Carlo (MCMC) sampling, we generated posterior distributions for $\eta_1,\eta_2,\eta_3,\eta_4,$ and  $\sigma_{\text{jitter}}$. Using the same GP kernel, we then modeled the observed RVs. The priors for $\eta_2,\eta_3,$ and $\eta_4$ were set by the posteriors of the activity index GP model. $\eta_1$ and $\sigma_{\text{jitter}}$ again had Jeffreys priors. GP models were implemented via \texttt{radvel} \citep{Fulton2018} and \texttt{emcee} \citep{ForemanMackey2013}.

\section{Model Comparison}
\label{sec:model_comparison}

As described in Section \ref{sec:mitigation}, we considered four primary methods of mitigating stellar activity: line selection, formation depth, FF$^{\prime}$, and GP regression. We conducted a systematic exploration of these four models using HARPS observations of $\alpha$CenB and HD~13808 and include linear decorrelation models for comparison.

\subsection{$\alpha$CenB}
\label{sec:alphaCenB}
We applied a series of activity mitigation methods to $\alpha$CenB HARPS observations from 2010. We constructed linear decorrelation, FF$^{\prime}$, and GP models using the depth metric, the FWHM of the CCF, and log$R^{\prime}_{HK}$. We also implemented two line-by-line based activity mitigation methods: line selection and formation depth. 

To ensure our mitigation methods are suppressing stellar activity while preserving Keplerian signals, we considered the unmodified RV time series (for which there is no significant planetary signal) and time series with injected planetary signals. Signals were injected by Doppler shifting the wavelength solutions of the stellar spectra in accordance with the injected planet's RV signature. We considered a five-day planet on a circular orbit for three injection scenarios: $K=0.75,1,$ and $2$~m s$^{-1}$. 

With the exception of the GP models, we considered the joint stellar activity and Keplerian model logarithmic likelihood function, 
\begin{align}
    \label{equ:likli_w_planets}
    \ln \mathcal{L}(  \text{RV}_j \mid \theta )  =& -\sum_{j=1}^{N} \ln \sqrt{2\pi (\sigma_{ \text{RV}_j}^2 + \sigma_{\text{jitter}}^2) } \nonumber\\ 
    &- \sum_{j=1}^{N} \frac{1}{2} \Bigg[ \frac{(\text{RV}_{j}-\text{RV}(t_j, \theta))^2 }{\sigma_{ \text{RV}_j}^2 + \sigma_{\text{jitter}}^2} \Bigg].
\end{align}
RV$(t_j, \theta)$ are the joint activity and Keplerian RV predictions, RV$_j$ are the observed values, $\sigma_{ \text{RV}_j}$ are the observational uncertainties, and $\sigma_{\text{jitter}}$ is an additive `jitter' noise term. For the GP kernels, we modeled the RVs using a joint activity-Keplerian GP model conditioned on an activity index (see Section \ref{sec:gp}). 

The Keplerian parameters of a given planet were $K$ (RV semi-amplitude), $P$ (orbital period), $e$ (eccentricity), $\omega$ (argument of periastron), and T$_0$ (time of transit). We adopted uniform priors of $0<K<10$~m~s$^{-1}$, $0.95 \cdot P_0<P<1.05 \cdot P_0$~days (for some central orbital period $P_0$), and $0<\text{T}_0<P$. We fixed $e=0$ and set $P_0=5$~days; for the null injection case ($K=0$~m s$^{-1}$), we froze the Keplerian component to $K=0$~m s$^{-1}$.

\startlongtable
\begin{deluxetable*}{lclccccc}
\tablecaption{ $\alpha$CenB activity mitigation and signal recovery\label{tab:RMS}}
\tablehead{
\colhead{Correction Method} & \colhead{RMS} & \colhead{Injection Case} &  \colhead{$K$} &  \colhead{$P$} & \colhead{ $\sigma_{\rm jitter}$}\\
\colhead{} & \colhead{(m s$^{-1}$)} & \colhead{} &  \colhead{(m s$^{-1}$)} &  \colhead{(days)} & \colhead{(m s$^{-1}$)}
}
\startdata
No correction (line-by-line RVs) & 2.67 &  --- & --- & --- & --- \\
No correction (DRS RVs) & 2.52 &  --- & --- & --- & --- \\
\hline
Line selection& 1.29  & --- &  --- & --- & --- \\
 & 1.25  & $K=0.75$ m s$^{-1}$& $0.495_{-0.206}^{+0.199} $ & $5.1084_{-0.0619}^{+0.0527} $ & $1.375_{-0.094}^{+0.103} $ \\
 & 1.26  & $K=1$ m s$^{-1}$& $0.677_{-0.207}^{+0.204} $ & $5.1083_{-0.0492}^{+0.0430} $ & $1.437_{-0.098}^{+0.110} $ \\
 & 1.16  & $K=2$ m s$^{-1}$& $1.281_{-0.179}^{+0.177} $ & $5.0390_{-0.0235}^{+0.0237} $ & $1.285_{-0.091}^{+0.095} $ \\
Formation depth& 1.59  & --- &  --- & --- & --- \\
 & 1.78  & $K=0.75$ m s$^{-1}$& $0.648_{-0.267}^{+0.258} $ & $5.1025_{-0.0539}^{+0.0489} $ & $1.654_{-0.126}^{+0.132} $ \\
 & 1.63  & $K=1$ m s$^{-1}$& $0.854_{-0.228}^{+0.229} $ & $5.0347_{-0.0337}^{+0.0385} $ & $1.450_{-0.117}^{+0.129} $ \\
 & 1.64  & $K=2$ m s$^{-1}$& $1.374_{-0.232}^{+0.230} $ & $5.0637_{-0.0293}^{+0.0299} $ & $1.500_{-0.116}^{+0.132} $ \\
 \hline
Linear decorrelation (log$R^{\prime}_{HK}$)& 1.67  & --- &  --- & --- & --- \\
 & 1.64  & $K=0.75$ m s$^{-1}$& $0.888_{-0.235}^{+0.231} $ & $5.0727_{-0.0425}^{+0.0472} $ & $1.682_{-0.109}^{+0.121} $ \\
 & 1.60  & $K=1$ m s$^{-1}$& $1.071_{-0.225}^{+0.217} $ & $5.0562_{-0.0351}^{+0.0370} $ & $1.637_{-0.104}^{+0.115} $ \\
 & 1.61  & $K=2$ m s$^{-1}$& $2.081_{-0.223}^{+0.228} $ & $5.0250_{-0.0174}^{+0.0181} $ & $1.649_{-0.105}^{+0.118} $ \\
Linear decorrelation (FWHM of the CCF)& 1.89  & --- &  --- & --- & --- \\
 & 1.82  & $K=0.75$ m s$^{-1}$& $0.970_{-0.248}^{+0.251} $ & $5.0505_{-0.0393}^{+0.0442} $ & $1.838_{-0.110}^{+0.098} $ \\
 & 1.69  & $K=1$ m s$^{-1}$& $1.165_{-0.235}^{+0.236} $ & $5.0410_{-0.0309}^{+0.0351} $ & $1.725_{-0.111}^{+0.123} $ \\
 & 1.68  & $K=2$ m s$^{-1}$& $2.175_{-0.231}^{+0.222} $ & $5.0202_{-0.0172}^{+0.0177} $ & $1.716_{-0.110}^{+0.115} $ \\
Linear decorrelation (depth metric)& 1.54  & --- &  --- & --- & --- \\
 & 1.49  & $K=0.75$ m s$^{-1}$& $0.922_{-0.206}^{+0.203} $ & $5.0529_{-0.0372}^{+0.0415} $ & $1.520_{-0.099}^{+0.113} $ \\
 & 1.47  & $K=1$ m s$^{-1}$& $1.123_{-0.207}^{+0.202} $ & $5.0438_{-0.0293}^{+0.0304} $ & $1.498_{-0.098}^{+0.108} $ \\
 & 1.53  & $K=2$ m s$^{-1}$& $2.075_{-0.212}^{+0.212} $ & $5.0193_{-0.0169}^{+0.0174} $ & $1.560_{-0.099}^{+0.114} $ \\
 \hline
FF$^{\prime}$ (log$R^{\prime}_{HK}$)& 1.40  & --- &  --- & --- & --- \\
 & 1.38  & $K=0.75$ m s$^{-1}$& $0.859_{-0.196}^{+0.213} $ & $5.0900_{-0.0429}^{+0.0419} $ & $1.419_{-0.097}^{+0.109} $ \\
 & 1.41  & $K=1$ m s$^{-1}$& $1.034_{-0.201}^{+0.214} $ & $5.0697_{-0.0332}^{+0.0342} $ & $1.458_{-0.099}^{+0.111} $ \\
 & 1.37  & $K=2$ m s$^{-1}$& $2.025_{-0.191}^{+0.197} $ & $5.0340_{-0.0168}^{+0.0162} $ & $1.414_{-0.092}^{+0.106} $ \\
FF$^{\prime}$ (FWHM of the CCF)& 1.37  & --- &  --- & --- & --- \\
 & 1.29  & $K=0.75$ m s$^{-1}$& $0.845_{-0.186}^{+0.184} $ & $5.0784_{-0.0416}^{+0.0417} $ & $1.333_{-0.096}^{+0.101} $ \\
 & 1.28  & $K=1$ m s$^{-1}$& $1.110_{-0.206}^{+0.193} $ & $5.0596_{-0.0294}^{+0.0286} $ & $1.324_{-0.087}^{+0.103} $ \\
 & 1.28  & $K=2$ m s$^{-1}$& $2.064_{-0.187}^{+0.183} $ & $5.0288_{-0.0150}^{+0.0162} $ & $1.324_{-0.088}^{+0.102} $ \\
FF$^{\prime}$ (depth metric)& 1.18  & --- &  --- & --- & --- \\
 & 1.20  & $K=0.75$ m s$^{-1}$& $0.848_{-0.169}^{+0.167} $ & $5.0521_{-0.0330}^{+0.0367} $ & $1.227_{-0.082}^{+0.091} $ \\
 & 1.22  & $K=1$ m s$^{-1}$& $1.044_{-0.166}^{+0.168} $ & $5.0467_{-0.0275}^{+0.0279} $ & $1.247_{-0.082}^{+0.091} $ \\
 & 1.23  & $K=2$ m s$^{-1}$& $1.996_{-0.181}^{+0.172} $ & $5.0216_{-0.0142}^{+0.0147} $ & $1.264_{-0.084}^{+0.097} $ \\
 \tablebreak
GP (log$R^{\prime}_{HK}$)& 1.02  & --- &  --- & --- & --- \\
 & 1.40  & $K=0.75$ m s$^{-1}$& $0.898_{-0.204}^{+0.183} $ & $5.0468_{-0.0007}^{+0.0002} $ & $1.048_{-0.089}^{+0.113} $ \\
 & 1.14  & $K=1$ m s$^{-1}$& $1.052_{-0.183}^{+0.188} $ & $5.0468_{-0.0003}^{+0.0001} $ & $1.048_{-0.095}^{+0.115} $ \\
 & 1.61  & $K=2$ m s$^{-1}$& $1.521_{-0.346}^{+0.364} $ & $5.0172_{-0.0001}^{+0.0001} $ & $1.074_{-0.106}^{+0.124} $ \\
GP (FWHM of the CCF)& 1.03  & --- &  --- & --- & --- \\
 & 1.39  & $K=0.75$ m s$^{-1}$& $0.900_{-0.182}^{+0.152} $ & $5.0487_{-0.0005}^{+0.0003} $ & $1.053_{-0.099}^{+0.103} $ \\
 & 1.06  & $K=1$ m s$^{-1}$& $1.037_{-0.168}^{+0.185} $ & $5.0486_{-0.0004}^{+0.0001} $ & $1.044_{-0.081}^{+0.106} $ \\
 & 2.22  & $K=2$ m s$^{-1}$& $1.693_{-0.433}^{+0.318} $ & $5.0172_{-0.0001}^{+0.0001} $ & $1.065_{-0.108}^{+0.120} $ \\
GP (depth metric)& 1.02  & --- &  --- & --- & --- \\
 & 1.23  & $K=0.75$ m s$^{-1}$& $0.891_{-0.214}^{+0.181} $ & $5.0468_{-0.0006}^{+0.0004} $ & $1.072_{-0.106}^{+0.109} $ \\
 & 0.97  & $K=1$ m s$^{-1}$& $1.076_{-0.184}^{+0.176} $ & $5.0468_{-0.0002}^{+0.0002} $ & $1.038_{-0.092}^{+0.105} $ \\
 & 1.45  & $K=2$ m s$^{-1}$& $1.654_{-0.489}^{+0.357} $ & $5.0171_{-0.0001}^{+0.0001} $ & $1.089_{-0.121}^{+0.132} $ \\
\enddata
\tablecomments{We report the 16th, 50th, and 84th percentiles of the posteriors. }
\end{deluxetable*}

For the stellar activity model components, both the line selection and formation depth methods had no free parameters, the FF$^{\prime}$ models each had six, the GP models had five, and the linear decorrelation models had two. For the line selection and formation depth methods, we adopted the spectral line subpopulations outlined in Sections \ref{sec:line_selection} and \ref{sec:formation_depth}, respectively. In principle, the thresholds used for defining these subpopulations could be treated as free parameters, but we leave such analysis to later work. 

The joint activity-Keplerian models were optimized via affine-invariant MCMC sampling using \texttt{emcee} \citep{ForemanMackey2013}. For the FF$^{\prime}$ and linear decorrelation models, we adopted uniform priors for the activity model parameters and employed 90 walkers for 10,000 iterations each, where the first 2,000 iterations were rejected as burn-in. For the GP models, we adopted the priors described in Section \ref{sec:gp} and employed 50 walkers for 10,000 iterations each.

\begin{deluxetable*}{lccc}[t]
\tablecaption{ HD~13808 two-planet model posteriors \label{tab:MAP}}
\tablehead{
\colhead{Correction Method} & & \colhead{HD~13808b} &  \colhead{HD~13808c}\\
}
\startdata
Line Selection &  &  & \\
 & K (m s$^{-1}$) & $2.700_{-0.370}^{+0.384}$ & $2.265_{-0.187}^{+0.299}$\\
 & P (days) & $14.184_{-0.004}^{+0.006}$ & $53.724_{-0.053}^{+0.069}$\\
 & $e$ & $0.292_{-0.141}^{+0.135}$ & $0.192_{-0.134}^{+0.198}$\\
 & $\omega$ (rad) & $2.898_{-0.522}^{+0.454}$ & $4.296_{-1.732}^{+1.018}$\\
 & T$_0$ (days) & $4.673_{-1.005}^{+1.852}$ & $40.434_{-4.450}^{+3.604}$\\
 & $M\sin(i)$ (M$_{\oplus}$) & $8.155_{-1.151}^{+1.108}$ & $10.850_{-0.866}^{+1.332}$\\
 & $\sigma_{\rm jitter}$ (m s$^{-1}$) & $2.845_{-0.19}^{+0.208}$ & ---\\
Formation Depth &  &  & \\
 & K (m s$^{-1}$) & $2.906_{-0.637}^{+1.050}$ & $3.003_{-0.650}^{+0.880}$\\
 & P (days) & $14.190_{-0.014}^{+0.023}$ & $53.762_{-0.086}^{+0.101}$\\
 & $e$ & $0.510_{-0.341}^{+0.295}$ & $0.358_{-0.237}^{+0.290}$\\
 & $\omega$ (rad) & $3.499_{-1.954}^{+1.294}$ & $4.278_{-1.576}^{+0.823}$\\
 & T$_0$ (days) & $7.456_{-4.569}^{+3.568}$ & $47.358_{-13.427}^{+4.377}$\\
 & $M\sin(i)$ (M$_{\oplus}$) & $7.531_{-1.763}^{+2.117}$ & $13.224_{-2.798}^{+3.318}$\\
 & $\sigma_{\rm jitter}$ (m s$^{-1}$) & $4.229_{-0.636}^{+0.633}$ & ---\\
FF$^{\prime}$ (depth metric) &  &  & \\
 & K (m s$^{-1}$) & $3.872_{-0.298}^{+0.301}$ & $2.552_{-0.268}^{+0.293}$\\
 & P (days) & $14.179_{-0.003}^{+0.003}$ & $53.869_{-0.059}^{+0.066}$\\
 & $e$ & $0.057_{-0.040}^{+0.061}$ & $0.122_{-0.086}^{+0.138}$\\
 & $\omega$ (rad) & $4.159_{-2.090}^{+1.320}$ & $2.120_{-1.263}^{+2.000}$\\
 & T$_0$ (days) & $6.718_{-0.573}^{+0.573}$ & $33.345_{-3.670}^{+3.371}$\\
 & $M\sin(i)$ (M$_{\oplus}$) & $12.288_{-0.953}^{+0.958}$ & $12.468_{-1.292}^{+1.372}$\\
 & $\sigma_{\rm jitter}$ (m s$^{-1}$) & $2.619_{-0.143}^{+0.154}$ & ---\\
 \hline
 \cite{Ahrer2021} &  &  & \\
 & K (m s$^{-1}$) & $3.67 \pm 0.22$ & $2.18^{+0.22}_{-0.20}$\\
 & P (days) & $14.1815 \pm 0.0015$ & $53.753^{+0.050}_{-0.082}$\\
 & $e$ & $0.071^{+0.027}_{-0.047}$ & $0.156^{+0.050}_{-0.061}$\\
 & $M\sin(i)$ (M$_{\oplus}$)  & $11.2^{+1.2}_{-0.66}$ & $9.96^{+1.8}_{-0.96}$\\
\enddata
\tablecomments{We report the 16th, 50th, and 84th percentiles of the posteriors. T$_0$ values are relative to BJD$=2452985.6179$. For comparison, we also present the maximum a posteriori (MAP) values from the preferred model of \cite{Ahrer2021}. }

\end{deluxetable*}

\begin{figure*}[t]
\gridline{\fig{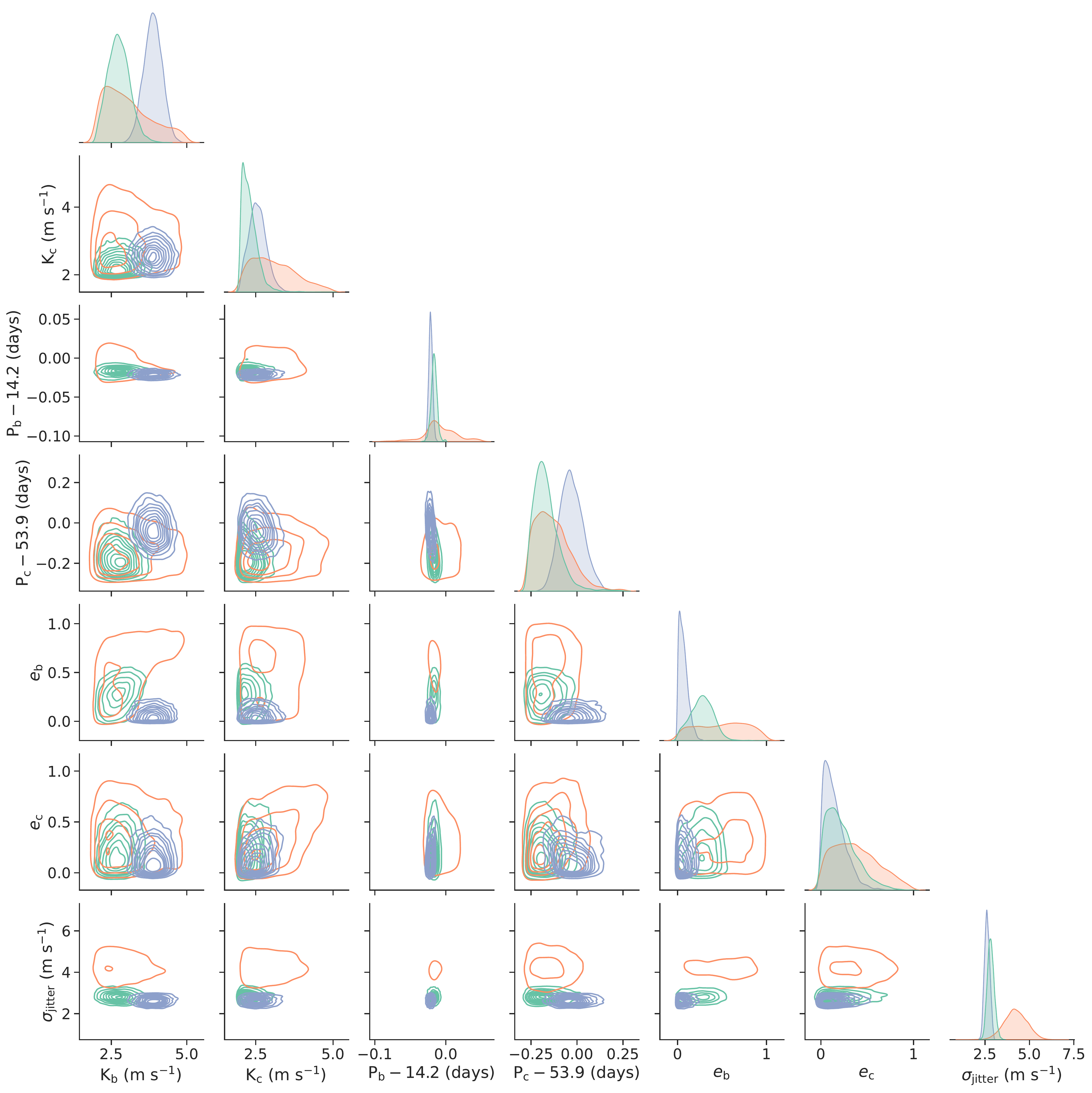}{.9\textwidth}{}}
\caption{The depth metric based FF$^{\prime}$ model yields well-constrained posteriors for HD~13808b and HD~13808c. We present the corner plot of the activity-corrected two-planet model posteriors for RV semi-amplitudes ($K$), orbital periods ($P$), eccentricities ($e$), and `jitter' ($\sigma_{\rm jitter}$). The posteriors for the line selection model are shown in green, the formation depth posteriors are shown in orange, and the depth metric based FF$^{\prime}$ posteriors are shown in blue.}

\label{fig:corner}
\end{figure*}

\begin{figure*}[t]
\gridline{\fig{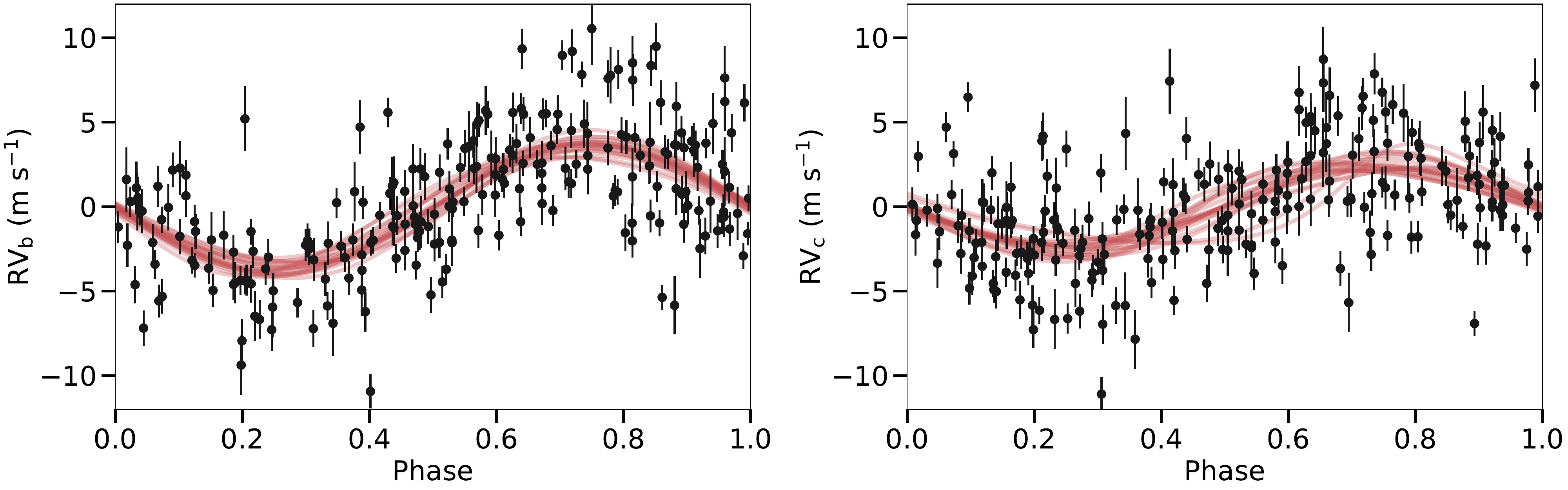}{\textwidth}{}}
\caption{Phase-folded RVs of HD~13808b (\textit{left}) and HD~13808c (\textit{right}) for the depth metric FF$^{\prime}$ two-planet model. For each planet, we present the observed RVs minus the stellar-activity correction and the RV model for the other planet (black circles). Random posterior samples for each planet's RV model are shown in red.}
\label{fig:phased_rvs}
\end{figure*}

For the null-injection case, the corrected RVs for a selection of mitigation methods are shown in Figure \ref{fig:alphaCenB_mitigation}. Each method shows a significant reduction in RV scatter and reduces the amplitude of the stellar activity signal. In terms of RMS, the FF$^{\prime}$ and GP models perform the best. For all methods, the resulting RMS is considerably greater than the RV uncertainties. The bulk line-by-line RV measurements have a mean binned (to one tenth of a day) RV uncertainty of $0.18$~m~s$^{-1}$. The line selection and formation depth methods have mean binned uncertainties of  $0.33$~m~s$^{-1}$ and $0.68$~m~s$^{-1}$, respectively. These uncertainties only consider the uncertainties in $A\delta\lambda$ and $A$ from the least-squares fits to each line. The DRS RVs have a mean binned RV uncertainty of $0.18$~m~s$^{-1}$.

To gauge the mitigation methods' planet detection capabilities, in Figure \ref{fig:alphaCenB_injection} we present the Lomb-Scargle periodograms of the activity-corrected RVs for a selection of mitigation methods under each injection case. To mimic a blind planet search, we omit the Keplerian component and model the RVs using pure activity models. Each correction method successfully recovers the $K=2$~m~s$^{-1}$ planet at a 1\% false-alarm probability level, a helpful but nonrigorous standard. However, only the FF$^{\prime}$ models significantly recover the $K=1$~m~s$^{-1}$ signal. While the depth metric based FF$^{\prime}$ model successfully recovers the $K=0.75$~m~s$^{-1}$ signal at a 1\% false-alarm probability level, the detection is only marginally better than the FWHM FF$^{\prime}$ model. Without a Keplerian component, the GP-based models had a strong tendency to absorb the injected planetary signals into the activity component of the model. This is less conducive to planet discovery than the other methods, which enhance the injected planetary signal after removing activity (Figure~\ref{fig:alphaCenB_injection}).

Table \ref{tab:RMS} shows the residual RMS and the recovered planet parameters for each combination of mitigation method, activity index, and injection scenario. In Figure \ref{fig:injection_recovery}, we compare the recovered RV semi-amplitude for each mitigation method. With the exception of the line selection and formation depth methods, we find every mitigation method recovers the semi-amplitude within approximately $1\sigma$ for each injection case and constrains the orbital period to within $0.1$~days. Between the GP and FF$^{\prime}$ models, the FF$^{\prime}$ models consistently yield lower uncertainties. 
 
The FF$^{\prime}$ models perform quite similarly across the three input activity indices. The depth metric FF$^{\prime}$ model yields marginally lower uncertainties, $\sigma_{\text{jitter}}$, and RMS than the FWHM- and log$R^{\prime}_{HK}$-based FF$^{\prime}$ models. For the GP models, we find no significant dependence on the input activity index. Unlike the FF$^{\prime}$ and linear decorrelation methods, the GP models do not directly detrend the RVs using an activity index. Instead, the GP models are conditioned on the activity index by adopting the posteriors of the activity index GP model as the priors of the RV model. 

With varying degrees of success, all mitigation methods considered here reduce the rotationally-modulated activity signal in the $\alpha$CenB time series. Evaluated on residual scatter and signal recovery, we favor the FF$^{\prime}$ and GP models over the line selection, formation depth, and linear decorrelation methods. For all activity models, the depth metric performs as well as, or marginally better than, the FWHM of the CCF and log$R^{\prime}_{HK}$. 

\subsection{Planet Characterization: HD~13808}

Having validated that our models successfully track rotationally-modulated stellar activity, we next applied these models to a system with a multiyear magnetic activity cycle. For this exploration, we considered HD~13808, which was recently studied with several stellar activity mitigation methods, including linear decorrelation, CCF FWHM-based FF$^{\prime}$, harmonic activity modeling, and Gaussian processes regression \citep{Ahrer2021}. That study confirmed the presence of two exoplanets, HD~13808b and HD~13808c, with orbital periods of 14.2 and 53.8 days and minimum masses of 11 and 10 M$_{\oplus}$. 

We employed joint activity-Keplerian FF$^{\prime}$, line selection, and formation depth models. These models only differ from those considered in Section \ref{sec:alphaCenB} by the addition of a second Keplerian component. Free parameters in the models were inferred via MCMC sampling with 90 walkers for 100,000 iterations each, where the first 20,000 iterations were rejected as burn-in. We adopted uniform priors of $2<K<10$~m s$^{-1}$, $0.95 \cdot P_0<P<1.05 \cdot P_0$~days (for some central orbital period $P_0$), $-3<\log_{10}(e)<0$, $0<\omega<2\pi$, and $0<\text{T}_0<P$. For HD~13808b and HD~13808c, we set $P_0$ following the periods reported by \cite{Ahrer2021}. 

Figure \ref{fig:corner} shows the posterior distributions of the line selection, formation depth, and depth metric based FF$^{\prime}$ two-planet models. The phase-folded RVs of HD~13808b and HD~13808c for the depth metric FF$^{\prime}$ two-planet model are shown in Figure \ref{fig:phased_rvs}. The 16th, 50th, and 84th percentiles for each model's posteriors are reported in Table \ref{tab:MAP}.  Both the line selection and FF$^{\prime}$ corrections achieve well-constrained posteriors and yield median parameter values in agreement with the preferred model of \cite{Ahrer2021}. On the other hand, the formation depth method struggles to characterize the HD~13808 system. In addition to favoring high stellar jitter, the formation depth model yields wide posteriors for the RV semi-amplitudes and favors significantly higher eccentricities. 

Although the line selection and depth metric based FF$^{\prime}$ models both yield well constrained posteriors, there are notable differences between the two. Compared to the line selection model, the FF$^{\prime}$ method yields considerably narrower eccentricity posteriors; the line selection model has a significant high eccentricity skew for both HD~13808b and HD~13808c. Moreover, the two models are rather discrepant for the RV semi-amplitude of HD~13808b and HD~13808c, with the line selection model favoring lower amplitude solutions. A similar trend was seen in Section \ref{sec:alphaCenB}. 

For a quantitative assessment of each models' relative evidence, we considered the deviance information criterion (DIC), the Bayesian information criterion (BIC), and the Watanabe-Akaike information criterion (WAIC). Following \cite{Gelman2013}, we found information criterions relative to the depth metric based FF$^{\prime}$ two-planet model of $\Delta$DIC$=$90 and 490, $\Delta$BIC$=$40 and 440, and $\Delta$WAIC$=$70 and 460 for the line selection and formation depth two-planet models, respectively. We thus found strong evidence that the formation depth two-planet model is insufficient, and therefore we favor the FF$^{\prime}$ two-planet model over the line selection model. However, greater exploration of all three methods is warranted, as there is room for optimization of the line selection and formation depth methods.

\section{Conclusions}
\label{sec:concl}

We introduced a high signal-to-noise stellar activity indicator, the depth metric $\mathcal{D}(t)$, which co-adds depth variations across a set of spectral lines shown to be activity-sensitive.  Through line-by-line radial velocity analysis of HARPS spectra, we identified $>60$ such lines for $\alpha$CenB and $>50$ lines for HD~13808. This novel activity indicator has the benefit of being self-contained, computationally inexpensive, and translation-invariant. Unlike chromospheric indices, such as log$R^{\prime}_{HK}$, this metric does not require coverage of a particular part of the spectrum. \cite{Wise2018} also postulated that spectral line depth variations are tied to Zeeman splitting from star spots, potentially making the depth metric a more direct probe of the stellar surface; however, the connection between spectral line distortions and physical processes within a host star remains poorly understood and warrants further study. 

We then conducted a comparative activity mitigation study. We considered linear decorrelation, FF$^{\prime}$, and GP regression models using three separate activity indices (the depth metric $\mathcal{D}(t)$, FWHM of the CCF, and log$R^{\prime}_{ HK}$), as well as two line-by-line based activity mitigation techniques, namely line selection \citep{Dumusque2018} and formation depth \citep{Cretignier2020}. We found the $\mathcal{D}(t)$-based FF$^{\prime}$ and GP models outperformed the linear decorrelation and line-by-line methods and reached the quality of the FWHM- and log$R^{\prime}_{HK}$-based FF$^{\prime}$ and GP models. The depth metric based FF$^{\prime}$ model successfully reduced the $\alpha$CenB RV RMS from 2.67 to 1.18~m~s$^{-1}$ and accurately characterized a five-day $K=1$ m~s$^{-1}$ injected signal; the depth metric based GP model yielded a RMS of 1.02~m~s$^{-1}$. 

Having proven it to be a powerful tool for mitigating the rotationally modulated activity signal of $\alpha$CenB, we next applied the depth metric to HARPS observations of HD~13808. This system presents a clear multiyear stellar activity cycle and hosts two confirmed exoplanets. Between the line selection, formation depth, and $\mathcal{D}(t)$ FF$^{\prime}$ models, only the line selection and FF$^{\prime}$ two-planet models yielded well-constrained posteriors for HD~13808b and HD~13808c; these models generally agree with the preferred model of \cite{Ahrer2021}. The deviance information criterion, the Bayesian information criterion, and the Watanabe-Akaike information criterion all significantly favor the depth metric based FF$^{\prime}$ model. 

We anticipate the depth metric will facilitate high-precision stellar activity tracking for a variety of applications. In the future, the depth metric has the potential for an exciting range of applications and elaborations. Below, we briefly describe several potential avenues for future research.

For Sun-like stars with activity signals on the m~s$^{-1}$ level, the depth metric has been shown to successfully track stellar activity with a level of quality similar to that provided by the FWHM of the CCF and log$R^{\prime}_{HK}$. Since the depth metric can be derived independently from these commonly used indices, it offers an additional dimension for detrending models and provides a valuable cross-check in cases of contentious detections; applying the depth metric alongside existing activity indices should yield more robust activity tracking and false-positive flagging. For stars with relatively weak activity signals, the co-adding of activity-sensitive spectral lines should instill the depth metric with higher signal-to-noise than existing indices, which may be dominated by white noise for low-activity stars; however, applying the depth metric to such stars may require modifications to the methods presented in Section \ref{sec:methods}, see below for discussion.

Presently, the depth metric method relies on the identification of activity-sensitive spectral lines through Pearson correlation coefficient analysis. However, this analysis relies on a long observing baseline and a moderate- to high-amplitude stellar activity signal; this is particularly true when planets are present. Since these restrictions do not apply to most commonly used activity indicators (e.g., FWHM of the CCF and log$R^{\prime}_{HK}$), addressing these limitations is key to making the depth metric a scalable and accessible activity indicator. To facilitate the application of the depth metric method to all stars, regardless of observational cadence or activity level, a catalog of lines shown to be sensitive to stellar activity for a given stellar type should be generated. Such a catalog could also extend into the far red using data from MAROON-X, NEID, ESPRESSO, and KPF. For a given star, a set of activity-sensitive spectral lines could be interpolated from such a catalog. Alternatively, if a star hosts a planetary system, activity-sensitive spectral lines could be identified by constraining the activity sensitivity of each spectral line simultaneously with the planets' parameters. 

The depth metric method and the broader \cite{Dumusque2018} and \cite{Cretignier2020} methods only consider isolated spectral lines. This poses a considerable challenge when applying the line-by-line technique to stellar types dominated by blended lines (e.g, M-dwarfs). Future modifications to the line-by-line method that consider blended lines would expand the stellar types accessible to line-by-line analysis.

Finally, solar line-by-line measurements, specifically $\mathcal{D}(t)$, should be mapped to independently derived solar observables that directly trace activity processes \citep{Haywood2020}. Such studies would help constrain the relation between magnetic fields and granulation on the stellar surface to RV variations.

\ \\
\acknowledgments
We thank Jacob Bean and Andreas Seifahrt for helpful conversations and for spectra used for experimentation with this technique.  We acknowledge NSF grant AST-2034278.  R.A.R.\ is supported by an NSF Graduate Research Fellowship, grant No.\ DGE 1745301.

\bibliography{paper}%

\end{document}